\documentclass[journal]{IEEEtai}
\usepackage[colorlinks,urlcolor=blue,linkcolor=blue,citecolor=blue]{hyperref}
\usepackage{amsmath,amssymb,amsfonts}
\usepackage{color,array}
\usepackage[T1]{fontenc}
\usepackage{graphicx}
\usepackage{booktabs, multirow, colortbl, xcolor}
\usepackage{orcidlink}
\usepackage{float}
\usepackage{tabularx}
 
\usepackage{threeparttable}
\usepackage{pifont}

\newcommand{\cmark}{\textcolor{green!60!black}{\ding{51}}}   
\newcommand{\pmark}{\textcolor{orange!85!black}{\ensuremath{\triangle}}} 
\newcommand{\xmark}{\textcolor{red}{\ding{55}}}    
\usepackage{multirow}

\begin{document}

\title{Adversarial Defense in Cybersecurity: A Systematic Review of GANs for Threat Detection and Mitigation} 

\author{Tharcisse Ndayipfukamiye\,\orcidlink{0000-0002-1325-9834},
        Jianguo Ding$^{*}$\,\orcidlink{0000-0002-8927-0968}, 
        Doreen Sebatian Sarwatt Ding$^{*}$\,\orcidlink{0000-0002-3828-432}, Adam Gaston Philippo$^{*}$\,\orcidlink{0009-0009-2997-2990}, and Huansheng Ning$^{*}$\,\orcidlink{0000-0001-6413-193X} 
        \thanks{$^{*}$Corresponding authors: J. Ding (jianguo.ding@bth.se) and H. Ning (ninghuansheng@ustb.edu.cn). Please address correspondence to both authors.}
        \thanks{Manuscript received Month XX, 2024; revised Month XX, 2025.}
\thanks{T. Ndayipfukamiye is with the Department of Computer Science and Technology, School of Computer and Communication Engineering, University of Science and Technology Beijing, 10083, Haidian Beijing, China (email: d202461029@xs.ustb.edu.cn.}
\thanks{J. Ding, is with the Department of Computer Science, Blekinge Institute of Technology, 37179, Karlskrona, Sweden (email: jianguo.ding@bth.se).}
\thanks{H. Ning is with the Department of Computer Science and Technology, School of Computer and Communication Engineering, University of Science and Technology Beijing, 10083, Haidian, Beijing, China (email: ninghuansheng@ustb.edu.cn).}
\thanks{This paragraph will include the Associate Editor who handled your paper.}}
\maketitle
\begin{abstract}
Generative adversarial networks (GANs) are increasingly used in cybersecurity, yet their defensive value remains difficult to compare across heterogeneous architectures, threat models, datasets, and evaluation protocols. We conducted a PRISMA~2020 systematic review of peer-reviewed literature published from January~2021 through June~2026 across six electronic sources. The final corpus comprised ($N=289$) included full-text publications, of which ($n=231$)
reported an identifiable original experimental evaluation and formed the
experiment-level evidence subset. The literature is dominated by anomaly and threat detection (69.2\%), while 67.8\% of publications use a general or unspecified primary threat condition. Evidence maturity remains limited: 45.3\% of publications are offline proof-of-concept studies and none demonstrates sustained operational validation. Within the experimental subset, comparators are common (89.6\%), but explicit ablation (13.4\%), repeated-run evaluation (6.9\%), and public author artifacts (7.4\%) remain uncommon. Architecture and defensive function show little systematic association
(\(V=0.141\), \(V_{\mathrm{bc}}=0.000\)), indicating that named GAN
families do not sharply determine defensive role. Defensive function and
threat/data condition are strongly aligned
(\(V=0.604\), \(V_{\mathrm{bc}}=0.592\)), although this magnitude
partly reflects construct-coupled coding; more substantively, 179/200
detection publications (89.5\%) use a general or unspecified primary
threat condition. The evidence therefore supports conditional, task-specific uses of GANs rather than universal architectural superiority, and highlights the need for threat-explicit, reproducible, and deployment-oriented evaluation. The review protocol snapshot, screening audit, coding registry, and supporting materials are openly archived at~\url{https://doi.org/10.17605/OSF.IO/WQHSP}
\end{abstract}

\maketitle
\IEEEpeerreviewmaketitle


\section{Introduction}
\label{sec:introduction}

AI-based cybersecurity faces a structural challenge: important attacks are often those for which reliable training data are least available. Intrusion detection, malware classification, fraud detection, and Internet of Things (IoT) security must operate with imbalanced, rapidly changing, sensitive, and potentially manipulated data. Rare attacks may be sparsely represented, previously unseen threats may differ from development data, and adaptive adversaries can exploit learning systems through evasion, poisoning, backdoors, and related attacks
~\cite{11012727,HALLAJI2023110384,10945280,electronics14030574,
RANDHAWA2024294,10574838,11408066}. Conventional security mechanisms remain essential, but rule- and history-dependent defenses alone cannot cover polymorphic, zero-day, and distribution-shifting threats~\cite{11478946,10604482}. Deep learning improves representation capability but remains sensitive to class scarcity, dataset artifacts, label noise, domain shift, and adversarial manipulation~\cite{ROSHAN2024103853,Dube2024,LAYEGHY2024103689,11393807}.

Generative Adversarial Networks (GANs) provide one response to these limitations by learning data distributions through adversarial interaction between a generator and discriminator~\cite{NIPS2014_f033ed80}. Conditional,
Wasserstein, and gradient-penalized variants improve controllability or training stability~\cite{DBLP:journals/corr/MirzaO14,pmlr-v70-arjovsky17a,NIPS2017_892c3b1c}. In cybersecurity, GANs now support several distinct defensive roles: rare-class augmentation~\cite{ZHAO2024104005,SRIVASTAVA2023103432}, adversarial hardening~\cite{9863068,QIN2024110655,AlEssa2026PEAdversarialGAN}, anomaly detection~\cite{BOPPANA2023105805,MISHRA2024493}, privacy-aware or federated collaboration~\cite{TABASSUM2022299,Feizi2024,Ullah2026}, and domain adaptation~\cite{BHARDWAJ2024103769,e23081089}. Applications further span biometrics, fraud, cyber--physical systems, malware, IoT, and autonomous environments~\cite{app12031454,s21030699,math11194160,9667381,YAO2023103622,KUMAR2023103054,10749874}.

This diversity complicates evidence synthesis. A GAN used for class balancing, a reconstruction-based anomaly detector, a federated generator, and an adversarial-training system address different security failures and therefore require different comparators, outcomes, threat assumptions, and validation conditions~\cite{Yao2025CWGANGP,10.1007/978-3-030-20893-6_39}. Favorable
benchmark scores may reflect class imbalance, leakage, weak baselines, or auxiliary pipeline components rather than the incremental value of the GAN itself. GANs are also dual use: mechanisms developed for defensive synthesis, testing, and hardening may support evasive malware, deceptive traffic, or adversarial generation~\cite{ZHU2022485,HOANG2024104031,
COPPOLINO2025129406}. A useful review must therefore move beyond cataloging architectures and accuracy values to determine when GAN-based defenses are supported by credible, reproducible, and operationally meaningful evidence.

\subsection{Motivation and Knowledge Gap}
\label{subsec:gap}

Existing reviews provide useful accounts of GAN applications in cybersecurity, intrusion detection, anomaly detection, and the dual use of generative models
~\cite{arifin2024surveyapplicationgenerativeadversarial,ALAJLAN20242053,Alauthman2026,LIM2024103733,SAEED2026128978,COPPOLINO2025129406}. However, the field has expanded beyond what an architecture- or application-centered survey can explain. Four gaps remain particularly consequential:
First, \textbf{Fragmented design logic.} Existing classifications commonly organize studies by GAN family, security application, or a single defensive function, but rarely connect defensive function, architecture, cybersecurity domain, threat or data condition, and evidence context within one design space. Architecture prevalence can therefore be mistaken for suitability, effectiveness, or maturity~\cite{YinkaBanjo2020,10187144,ALAJLAN20242053,LIM2024103733,arifin2024surveyapplicationgenerativeadversarial,Alauthman2026}.
Second, \textbf{Weak comparability of reported effectiveness.} Accuracy, macro-F1, AUC, attack-success reduction, privacy leakage, synthetic-data fidelity, latency, and energy represent substantively different outcomes. Comparability is further weakened by differences in datasets, class prevalence, preprocessing, train--test partitions, baselines, thresholds, attacker knowledge, and reporting practices~\cite{ALAJLAN20242053,LIM2024103733,SAEED2026128978,Dube2024,LAYEGHY2024103689,11393807}.
Third, \textbf{Limited attention to evidence credibility.} The incremental value of a GAN depends on matched ablations, strong function-specific comparators, leakage control, repeated runs, uncertainty estimates, transparent configurations, and reproducible artifacts. Although recent reviews discuss methodological challenges, these factors are not consistently integrated into GAN classifications or used to qualify review conclusions~\cite{BOUNTAKAS2023100573,10584534,ALAJLAN20242053,Alauthman2026,doi:10.1126/science.359.6377.725,10778911}.
Fourth, \textbf{A gap between benchmark feasibility and deployment readiness.} Most studies emphasize offline predictive performance, whereas external and temporal generalization, adaptive attackers, end-to-end latency, energy, communication cost, privacy, analyst workload, maintenance, and rollback are
evaluated far less consistently~\cite{s26030757,10430114,11037504}.

These gaps indicate that the principal issue is no longer simply whether GANs can improve cybersecurity models. The review is therefore guided by the following overarching question:

\begin{quote}
\emph{Under which data, threat, architectural, and deployment conditions do GANs provide credible value for cybersecurity defense, and how trustworthy is the supporting evidence?}
\end{quote}

Answering this question requires an evidence-centered synthesis that treats the GAN architecture as one element of a broader defensive system and keeps retrieved records, included primary studies, prior reviews, offensive-context studies, and foundational references analytically distinct.

\subsection{Scope, Research Questions, and Contributions}
\label{subsec:research_questions}

As detailed in Section~\ref{sec:methodology}, this study covers peer-reviewed publications published from January~2021 through June~2026 in which a GAN or GAN-derived architecture makes a direct and substantive contribution to cybersecurity
defense, detection, mitigation, protection, privacy-aware collaboration,
adaptation, deception, testing, or recovery. Original experimental status is
determined only after review-corpus inclusion and is used to define the
experiment-level analytical subset.

To address the gaps identified in Section~\ref{subsec:gap}, this review is guided by five research questions (RQs):

\begin{itemize}
\item \textbf{RQ1 (Evidence landscape):}
How has GAN-based cybersecurity defense evolved in terms of defensive
functions, architectures, application domains, datasets, threat conditions,
and evaluation practices?
\item \textbf{RQ2 (Defensive design space):}
How can GAN-based defenses be classified by defensive function, architecture,
cybersecurity domain, threat or data condition, and evidence maturity?
\item \textbf{RQ3 (Empirical effectiveness):}
Under which datasets, comparators, metrics, threat assumptions, and evaluation
protocols do GAN-based defenses demonstrate credible benefits, and where is the
evidence insufficient?
\item \textbf{RQ4 (Evidence credibility and deployment readiness):}
What methodological, reproducibility, trustworthiness, and operational
limitations restrict current GAN-based cybersecurity defenses?
\item \textbf{RQ5 (Research and engineering priorities):}
What advances are required for more trustworthy and operationally credible
GAN-based cybersecurity defense?
\end{itemize}

Together, RQ1--RQ5 operationalize the overarching review question introduced in Section~\ref{subsec:gap}. The five questions therefore form a cumulative analytical sequence rather than independent lines of inquiry. In answering these RQs, this review makes five principal contributions:

\begin{enumerate}

\item \textbf{A traceable systematic evidence base.}
A PRISMA~2020 workflow yields \(N=289\) included full-text publications,
including \(n=231\) original experimental studies and 58 other included
publications retained for corpus-level mapping and qualitative synthesis.
The supplementary registry preserves the record-level audit trail and
evidence supporting taxonomy and experiment-level appraisal decisions.

\item \textbf{A five-dimensional taxonomy and cross-dimensional design space.}
The framework connects defensive function, GAN architecture, cybersecurity
domain, threat or data condition, and E1--E5 evidence maturity using one
primary category per dimension and secondary tags for hybrid characteristics.

\item \textbf{An outcome-specific, evidence-bounded empirical synthesis.}
Experiment-level synthesis is restricted to the \(n=231\) original
experimental studies. None of 150 candidate cross-study contrasts satisfied
the predefined split-and-preprocessing equivalence requirements, and six of
the 150 were invalid self-contrasts; pooled effects and directional vote
counting are therefore avoided.

\item \textbf{An integrated credibility and deployment analysis.}
The review links threat-model adequacy, comparator and ablation quality,
uncertainty, reproducibility, implementation transparency, resource cost,
privacy, and evidence maturity to the interpretation of reported defensive
performance.

\item \textbf{Actionable design guidance and a prioritized research agenda.}
The synthesis identifies the conditions under which GAN adoption is
defensible and the benchmark, methodological, reproducibility, and deployment
evidence required to strengthen future claims.
\end{enumerate}

Together, these contributions reposition GAN-based cybersecurity defense from a catalogue of architectures and benchmark scores to a problem of defensive design, evidence credibility, reproducibility, and operational fit.

\subsection{Paper Organization}

The remainder of this paper is organized as follows:
Section~\ref{sec:background_related} establishes the technical foundations and positions the review against prior surveys. Section~\ref{sec:methodology} reports the systematic protocol and evidence rules. Section~\ref{sec:taxonomy} presents the D1--D5 distributions, complete corpus traceability, cross-dimensional analysis, and quality findings. Section~\ref{sec:empirical_synthesis} provides the outcome-specific structured synthesis; Section~\ref{sec:challenges} examines technical, reproducibility, and deployment barriers; Section~\ref{sec:discussion_agenda} develops implications and research priorities; and Section~\ref{sec:conclusion} concludes.


\section{Conceptual Foundations and Related Reviews}
\label{sec:background_related}

This section provides only the conceptual material required to interpret the review. GAN-based cybersecurity defense is treated as a claim about an implemented security function, not merely as evidence that a generator can reproduce a data distribution. The section therefore links architectural choice to defensive purpose, defines the evidence needed to support security claims, and positions the review against adjacent syntheses.

\subsection{GAN Mechanisms as Components of Cybersecurity Defense}
\label{subsec:defensive_gan_foundations}

A GAN learns through competition between a generator $G$, which maps latent samples $z\sim p_z$ into the data space, and a discriminator $D$, which distinguishes generated samples from observations drawn from $p_{\mathrm{data}}$:
\begin{equation}
\min_G\max_D V(D,G)=
\mathbb{E}_{x\sim p_{\mathrm{data}}}[\log D(x)]+
\mathbb{E}_{z\sim p_z}[\log(1-D(G(z)))].
\label{eq:gan_objective}
\end{equation}
Conditional generation, Wasserstein objectives, and gradient-penalized training modify control or optimization, but they do not by themselves establish defensive effectiveness~\cite{NIPS2014_f033ed80,DBLP:journals/corr/MirzaO14,pmlr-v70-arjovsky17a,NIPS2017_892c3b1c}. In cybersecurity, statistical resemblance is insufficient: generated flows, executable representations, industrial sequences, authentication traces, or communication graphs must preserve protocol, temporal, behavioral, relational, or physical constraints~\cite{9877919,ROSHAN2024103853,Khalil2026}. A sample may appear realistic while being functionally impossible, or appear imperfect globally while still exposing a detector to a security-relevant decision boundary.

The defensive role of a GAN can therefore be reduced to four mechanisms. First, it can alter the training distribution through minority-class or rare-event synthesis. Second, it can model normality or reconstruct inputs to obtain anomaly scores. Third, it can generate attacks or perturbations for testing and hardening. Fourth, it can support distribution alignment, decentralized collaboration, or privacy-aware data exchange. These mechanisms are dual-use: the same capacity can synthesize evasive malware, deceptive traffic, or adversarial wireless signals~\cite{ZHU2022485,HOANG2024104031,COPPOLINO2025129406}. Accordingly, attack-generation papers enter the primary corpus only when they implement and evaluate a defensive, detection, mitigation, protection, or recovery function.

\subsection{From Architecture Choice to Claim Validity}
\label{subsec:architecture_design_logic}

Architecture is meaningful only in relation to the security failure it is intended to address. Conditional GANs permit class- or context-controlled synthesis; Wasserstein variants target optimization stability; reconstruction-oriented models support anomaly scoring; temporal and graph models represent sequential or relational dependence; and federated or privacy-oriented designs alter where data and model updates are exposed~\cite{NEURIPS2019_c9efe5f2,Wang_Wang_Wang_Zhao_Zhang_Zhang_Xie_Guo_2018,TABASSUM2022299}. None of these properties guarantees improved detection, robustness, privacy, or deployability. Hybrid systems further shift the estimand from the generator to the complete pipeline because feature selection, classifiers, autoencoders, transformers, explainability modules, and optimizers may account for the observed gain~\cite{MELISARAHEBI2025103745,HENNA2026114911,11250986}.

We therefore apply a claim--evidence alignment rule throughout the review. Augmentation claims require an unchanged downstream detector, conventional imbalance baselines, and untouched real test data. Hardening claims require an explicit attacker, budget, knowledge condition, clean-utility result, and preferably an adaptive or unseen attack. Zero-day, drift, and cross-domain claims require temporal, entity-, family-, device-, or domain-separated evaluation rather than random record-level splits. Privacy claims require direct leakage tests or formal guarantees; federation alone is not privacy evidence. Lightweight or real-time claims require end-to-end latency, throughput, memory, energy, and update costs on representative hardware~\cite{Feizi2024,10437148,s26030757,11037504}.

This rule also clarifies why the outcomes cannot be collapsed into a universal score. Accuracy, macro-F1, minority recall, false-positive burden, attack-success reduction, fidelity, memorization, privacy leakage, calibration, latency, energy, and throughput measure different properties. Credible synthesis must preserve the relation among defensive function, threat condition, data partition, comparator, outcome, and deployment boundary.

\subsection{Related Reviews and Remaining Gap}
\label{subsec:related_reviews_positioning}

Existing syntheses fall into three overlapping groups. General GAN reviews examine architectures, training challenges, regularization, and application breadth~\cite{10.1145/3463475,9625798,10.1145/3569928};
adversarial-machine-learning and broad AI-cybersecurity reviews organize
attacks, defenses, explainability, malware, and intrusion detection across multiple model families~\cite{10.1145/3453158,10.1145/3485133,
BOUNTAKAS2023100573,10073040,10926941}; and GAN-specific cybersecurity
reviews focus on areas such as intrusion and anomaly detection, malware,
IoT security, synthetic attack data, and dual use~\cite{YinkaBanjo2020,10187144,ALAJLAN20242053,LIM2024103733,
electronics13020322,10870477,11012727,Alauthman2026,SAEED2026128978}.
These studies establish the breadth of GAN use in cybersecurity, but generally treat defensive purpose, architecture, threat condition, experimental credibility, and evidence maturity separately or within narrower application domains. Table~\ref{tab:comparison} positions these features against representative prior reviews.

\begin{table*}[htbp]
\centering
\caption{Comparison of methodological and analytical characteristics across representative reviews of adversarial machine learning, AI-enabled cybersecurity, and GAN-based defense.}
\label{tab:comparison}
\scriptsize
\setlength{\tabcolsep}{2.0pt}
\renewcommand{\arraystretch}{0.5}
\begin{tabular}{lccccccp{7.5cm}}
\toprule
& \multicolumn{5}{c}{\textbf{Contributions}} & \multicolumn{1}{c}{\textbf{Scope}} & \multirow{2}{*}{\shortstack{\textbf{Primary focus and limitation}}}\\
\cmidrule(lr){2-6}
\cmidrule(lr){7-7}
\textbf{Study} & \textbf{\shortstack{Systematic\\Protocol}} &
\textbf{\shortstack{Structured\\Taxonomy}} & \textbf{\shortstack{Quantitative\\Synthesis}} &
\textbf{\shortstack{Quality /\\Reproducibility}} &
\textbf{\shortstack{Deployment\\Analysis}} & \textbf{\shortstack{GAN Focus}} \\
\midrule

\cite{10.1145/3453158} & \pmark & \cmark & \xmark & \xmark & \xmark & \xmark & Broad AML landscape; GANs peripheral.
\\
\cite{10.1145/3485133} & \pmark & \cmark & \xmark & \xmark & \xmark & \xmark & Defender-oriented AML taxonomy; mainly image classification.
\\
\cite{9693527} & \pmark & \cmark & \xmark & \xmark & \xmark & \xmark & AML in text processing; modality-specific and not GAN-centered.
\\
\cite{BOUNTAKAS2023100573} & \cmark & \cmark & \xmark & \cmark & \pmark & \xmark & Systematic AML-defense taxonomy; GANs treated as one family.
\\
\cite{10005100} & \pmark & \cmark & \xmark & \pmark & \pmark & \xmark & AML for NIDS; no cross-domain GAN-defense synthesis.
\\
\cite{10584534} & \cmark & \cmark & \xmark & \pmark & \pmark & \xmark & AML attacks and defenses; broader than GAN-based defense.
\\
\cite{Pelekis2025} & \pmark & \cmark & \xmark & \pmark & \pmark & \xmark & AML methods, tools, and sectors; GAN defenses not isolated.
\\
\midrule
\cite{10073040} & \cmark & \pmark & \xmark & \xmark & \pmark & \xmark & AI in cybersecurity; limited GAN-specific analysis.
\\
\cite{Salem2024} & \pmark & \pmark & \xmark & \xmark & \pmark & \xmark & AI-driven threat detection; GANs considered among many methods.
\\
\midrule
\cite{ALAJLAN20242053} & \xmark & \cmark & \xmark & \xmark & \pmark & \cmark & GAN architectures for IDS; no cross-domain synthesis. \\
\cite{LIM2024103733} & \cmark & \cmark & \pmark & \pmark & \cmark & \cmark & GAN-based network anomaly detection; domain-specific coverage.\\
\cite{arifin2024surveyapplicationgenerativeadversarial} & \xmark & \pmark & \xmark & \xmark & \pmark & \cmark & Broad GAN-cybersecurity overview; non-systematic synthesis. \\
\cite{COPPOLINO2025129406} & \xmark & \pmark & \xmark & \xmark & \pmark & \cmark & GAN attack--defense dual use; corpora not analytically separated.\\
\cite{Alauthman2026} & \cmark & \cmark & \xmark & \pmark & \pmark & \cmark & Systematic GAN-IDS review; limited cross-domain deployment analysis.\\
\cite{SAEED2026128978} & \pmark & \cmark & \xmark & \pmark & \pmark & \cmark & GAN-enabled anomaly detection; no formal effect synthesis.
\\
\midrule
\textbf{Present review} & \textbf{\cmark} & \textbf{\cmark} & \textbf{\pmark} &
\textbf{\cmark} & \textbf{\cmark} & \textbf{\cmark} & \textbf{Multi-domain GAN-defense review; structured SWiM synthesis without pooled effect estimation.}\\
\bottomrule
\end{tabular}
\vspace{1.5mm}
\footnotesize\textit{Note}: \cmark{} = criterion explicitly and substantially implemented; \pmark{} = criterion addressed only partially or within a restricted scope; \xmark{} = criterion not explicitly implemented or not central to the review. Ratings describe review characteristics rather than overall review quality. \textit{Systematic method} requires an explicit search, eligibility, and study-selection procedure; \textit{structured taxonomy} requires an explicit classification framework applied across the reviewed corpus; \textit{quantitative
synthesis} requires numerical cross-study synthesis beyond simple study listing; \textit{quality/reproducibility} requires structured study-level appraisal or reproducibility assessment; \textit{operational/deployment} requires explicit analysis of validation maturity or deployment evidence; and \textit{GAN-focused} indicates that GANs are the principal analytical unit. For the present review, quantitative synthesis is rated partial because no candidate contrast met the predefined requirements for pooled effectiveness estimation.
\end{table*}

The present review integrates these dimensions at study level and, more
importantly, uses them to quantify whether defensive claims are supported by threat-explicit, reproducible, and operationally mature evidence. Across the \(N=289\) included publications, 179/200 detection publications (89.5\%) use a general or unspecified primary threat condition, 131/289 (45.3\%) remain at E1 offline proof-of-concept, and none reaches E5 sustained operational validation. The \(n=231\) original experimental studies are further examined for comparators, ablations, uncertainty, reproducibility, and comparative admissibility. 

The unresolved problem is therefore one of integration and evidence. A defensible review must show what function a GAN performs (D1), which architectural mechanism it uses (D2), where it is deployed (D3), which threat or data condition motivates it (D4), and what level of evidence supports the claim (D5). It must then examine whether reported experiments isolate the GAN contribution, use security-valid partitions and comparators, quantify uncertainty, expose reproducible artifacts, and approach operational conditions. Sections~\ref{sec:taxonomy} and~\ref{sec:empirical_synthesis} provide that integrated analysis.


\section{Methodology}
\label{sec:methodology}

This review was designed as a systematic evidence map with structured quantitative synthesis. Reporting follows PRISMA~2020~\cite{PAGE2021103}; review planning, study selection, and extraction follow established computing and software-engineering guidance~\cite{KITCHENHAM2010792,10926941}; and quantitative synthesis without meta-analysis follows SWiM principles~\cite{Campbell2020SWiM}. A single record-level registry linked identification, query validation, deduplication, screening, retrieval, eligibility, extraction, D1--D5 coding, quality appraisal, and figure data.

\subsection{Review Scope, Questions, and Evidence Sets}
\label{subsec:review_design}

The five research questions defined in
Section~\ref{subsec:research_questions} are addressed through three evidence
layers. The \textbf{included review corpus} comprises \(N=289\) publications,
all examined in complete full text and used for corpus-level characterization,
D1--D5 mapping, cross-tabulation, and qualitative synthesis where applicable.

Within this corpus, \(n=231\) publications report an identifiable original
experimental evaluation and form the \textbf{original experimental subset}
used for analyses requiring experiment-level evidence. The remaining 58
publications remain fully included in the review but are outside
experiment-level frequency denominators; experiment-specific fields for these
publications are treated as not applicable rather than as missing or negative
evidence. A separate \textbf{contextual reference set} contains foundational,
methodological, benchmark, standards, comparator, and other sources used for
interpretation but excluded from review-corpus counts.

Thus,
\[
N
=
n_{\mathrm{experimental}}
+
n_{\mathrm{other\ included}}
=
231+58
=
289.
\]

The \(231/58\) distinction is a post-inclusion analytical stratification, not
an eligibility, retrieval, or full-text-access distinction. Corpus-level
analyses use \(N=289\), whereas experiment-level analyses use \(n=231\).
Sensitivity to this stratification is examined in
Section~\ref{subsec:sensitivity_limited_evidence}.

\subsection{Information Sources, Search Construction, and Search Amendments}
\label{subsec:search_strategy}

IEEE Xplore, ACM Digital Library, ScienceDirect, MDPI, SpringerLink, and
Scopus were searched for peer-reviewed English-language journal articles,
conference papers, and book chapters published from January~1, 2021 through
June~30, 2026. Search concepts combined GAN-family terms with cybersecurity
settings and defensive functions. Source-specific syntax, filters, retained
counts, export information, and provenance are reported in
Appendix~\ref{app:search_strings} and Supplementary Workbook~S1
~\cite{Ndayipfukamiye2026_GANProtocol}. Exact execution dates unavailable
from retained legacy exports were not reconstructed retrospectively.

\begin{figure}[!ht]
\centering
\includegraphics[width=\linewidth]{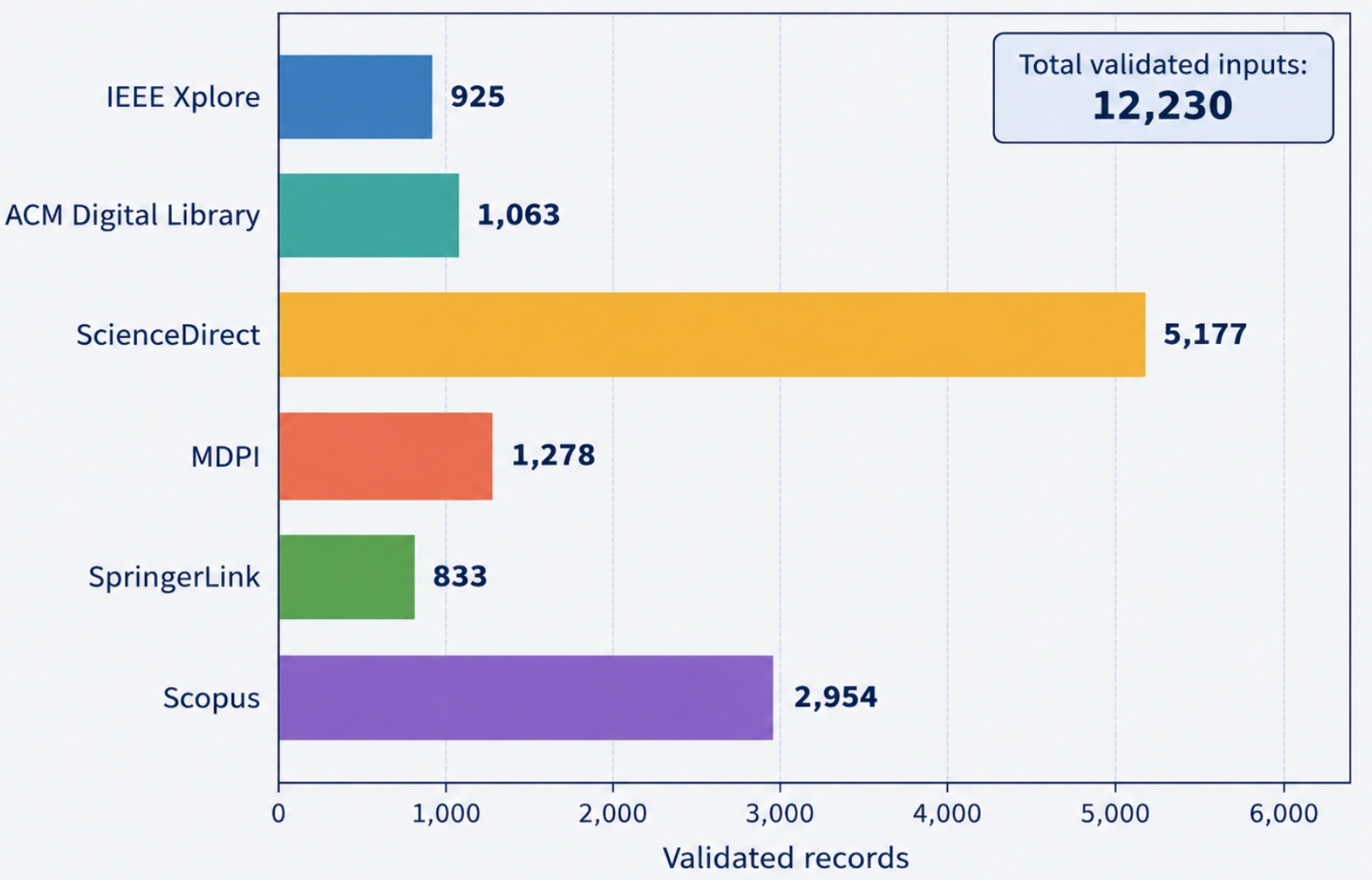}
\caption{\textbf{Validated records entering cross-source reconciliation.}
The six electronic sources contributed 12,230 validated pre-deduplication
records: IEEE Xplore (925), ACM Digital Library (1,063), ScienceDirect
(5,177), MDPI (1,278), SpringerLink (833 after removal of 167 documented
query/scope artifacts), and Scopus (2,954). The PRISMA identification total
retains all 12,397 records initially captured and reports the 167
SpringerLink removals separately.}
\label{fig:search_source_distribution}
\end{figure}

The searches captured 12,397 records: IEEE Xplore (925), ACM Digital Library (1,063), ScienceDirect (5,177), MDPI (1,278), SpringerLink (1,000), and
Scopus (2,954). SpringerLink query displayed 3,314 results, but the retained platform export/capture contained 1,000 records; after removal of 167 documented
query/scope artifacts, 833 entered cross-source reconciliation, and leaving 12,230 records for cross-source reconciliation. The SpringerLink pathway should therefore be regarded as export-limited rather than an exhaustive capture of all displayed results.Figure~\ref{fig:search_source_distribution} reports these validated inputs, including 833 SpringerLink records, whereas PRISMA retains the 12,397-record initial identification total and reports the 167 removals separately.

Scopus expansion and SpringerLink query validation were revision-stage protocol
amendments and are not presented as prospective. DOI normalization was the
primary matching rule, with normalized title and year used when DOI was
unavailable. Backward-reference reconciliation identified 32 candidates;
nine duplicated database records and 23 unique reports were retrieved and
included. Systematic forward-citation searching was not performed and is
treated as a coverage limitation. 

\subsection{Eligibility Criteria}
\label{subsec:eligibility}

Eligibility was determined at two levels:
\textit{review-corpus inclusion} and \textit{experiment-level inclusion}.

\subsubsection{Review-corpus inclusion.}
A publication was eligible when it:
(i) was peer reviewed and formally published between January~1, 2021 and
June~30, 2026;
(ii) addressed a GAN or GAN-derived architecture as a substantive component
of cybersecurity defense, detection, mitigation, protection, privacy-aware
collaboration, adaptation, deception, testing, recovery, or security
evaluation;
(iii) contained sufficient full-text information for at least one review
objective; and
(iv) was available in complete full text for eligibility assessment.
Original experimentation was not required for corpus inclusion, but
literature-level reviews, surveys, bibliometric studies, and similar secondary
syntheses were excluded from the \(N=289\) corpus.

\subsubsection{Experiment-level inclusion.}
After corpus inclusion, a publication entered the original experimental subset
when it reported:
(i) an implemented GAN or GAN-derived cybersecurity method;
(ii) an identifiable experimental or evaluation procedure;
(iii) an empirical dataset, simulation, testbed, or equivalent evaluation
setting; and
(iv) at least one measurable evaluation outcome.
These criteria yielded \(n=231\) original experimental studies; the remaining
58 publications remained included but outside experiment-level denominators.

\subsubsection{Exclusion criteria.}
Reports were excluded when they were outside the review window, were not peer
reviewed, fell outside cybersecurity, mentioned GANs only incidentally, lacked
a substantive GAN-based security role, focused exclusively on offensive
generation without a defensively relevant contribution, or were
literature-level secondary syntheses. Foundational papers, benchmark studies,
methodological guidance, standards, prior reviews, and other supporting
sources could be retained as contextual references. For 2026 publications,
eligibility followed the first peer-reviewed Version of Record or formal
publication date rather than preprint availability.

\subsection{Deduplication, Screening, Retrieval, and Eligibility Assessment}
\label{subsec:study_selection}

Screening followed a primary reviewer--verifier workflow. A second reviewer
independently reassessed a reproducible simple random sample of 62/308
full-text-assessed database/register reports (20.1\%). Agreement was 60/62
(96.8\%), with Cohen's \(\kappa=0.86\) (95\% CI [0.66, 1.00]); the two
disagreements were reconciled against the codebook. Because inclusion
prevalence was high, \(\kappa\) is interpreted alongside observed agreement.
The remaining records underwent primary-reviewer assessment with verifier
checks; complete independent dual screening is not claimed.

Of 10,734 title--abstract records, 10,016 were excluded and 718 reports were
sought. Of these, 410 (57.1\%) were not retrieved; 308 were assessed in
complete full text, yielding 42 exclusions and 266 database-pathway
inclusions. Backward-reference reconciliation added 23 unique full-text
publications, producing the final \(N=289\) corpus. After inclusion,
\(n=231\) publications formed the experimental subset and 58 remained in
corpus-level mapping and qualitative synthesis only.

All 410 non-retrieved reports originated from Scopus (410/624; 65.7\%).
Annual non-retrieval counts for 2021--2026 were 27, 34, 53, 74, 132, and 90,
respectively. Retrieval records comprised a composite
metadata/paywall/request-copy code for 323 reports, unavailable record-specific
cause for 34, metadata/preview-only access for 25, no complete report located
for 22, request-copy-only access for five, and pay-per-view plus request-copy
access for one. Their eligibility remains unknown; no content-based exclusion
or extrapolated inclusion count was assigned. The report-level audit is
archived in the OSF record~\cite{Ndayipfukamiye2026_GANProtocol}.

\begin{figure}[!ht]
\centering
\includegraphics[width=\linewidth]{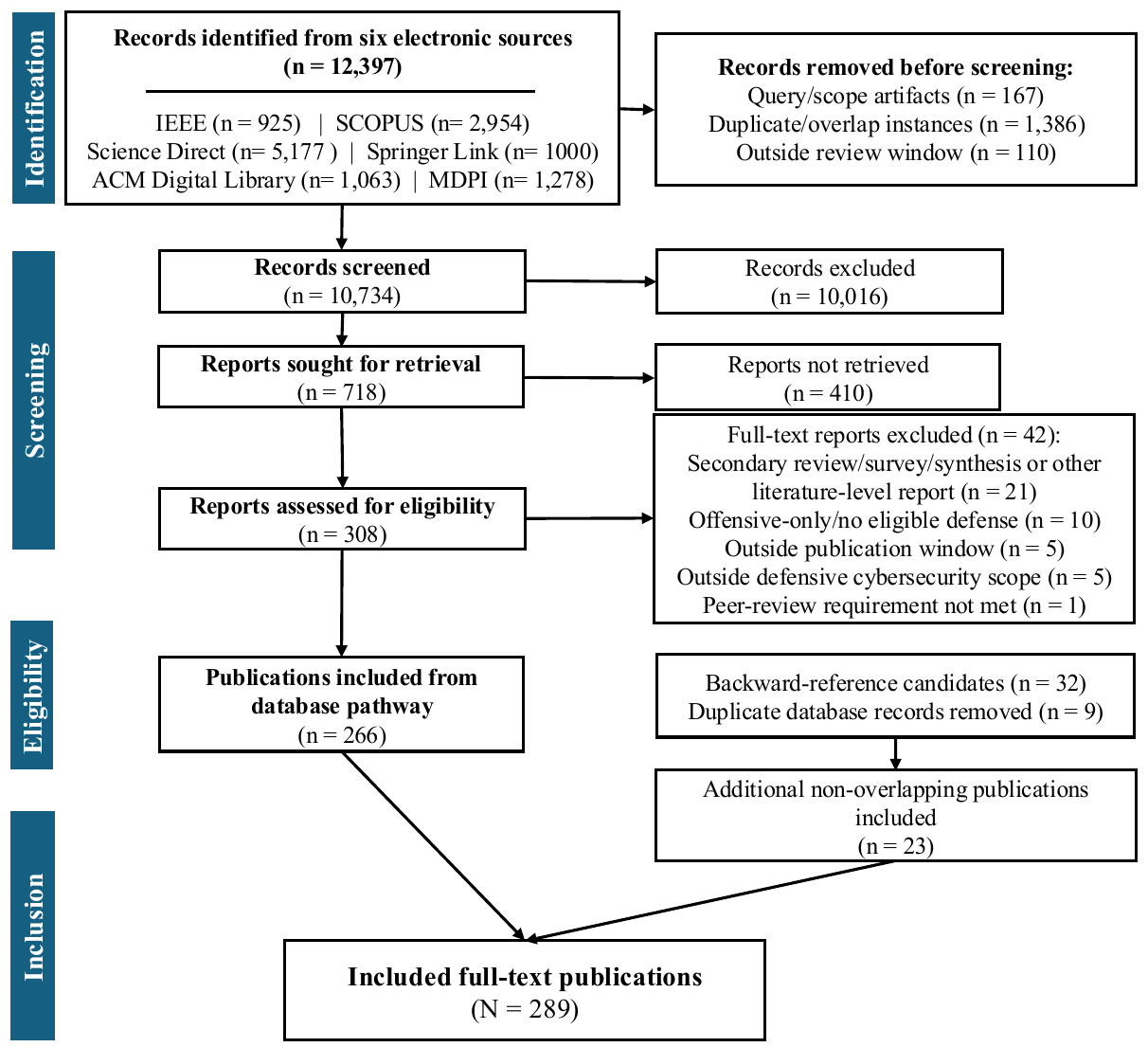}
\caption{\textbf{PRISMA 2020 study-selection flow.}
Database/register and backward-reference pathways are reported separately. The final included corpus comprises \(N=289\) publications, all examined in complete full text. After inclusion, the corpus was analytically stratified into \(n=231\) publications reporting an identifiable original experimental evaluation and 58 other included full-text publications retained for systematic mapping and qualitative synthesis but excluded from experiment-level frequency calculations. Contextual references are not included in these counts.}
\label{fig:prisma_flow}
\end{figure}

\subsection{Characteristics of Included Studies}
\label{subsec:included_characteristics}

Table~\ref{tab:database_year_distribution} summarizes the \(N=289\) included
publications by year and mutually exclusive primary discovery pathway.
Primary discovery source denotes where a publication entered the review
registry rather than every database indexing it; this prevents cross-source
double counting and reconciles with Figure~\ref{fig:prisma_flow}. The temporal
distribution is visualized in Figure~\ref{fig:temporal_distribution}.

\begin{table}[htbp]
\centering
\caption{Distribution of the \(N=289\) included full-text publications by primary
discovery source and publication year (2021--June 2026).}
\label{tab:database_year_distribution}
\scriptsize
\setlength{\tabcolsep}{2pt}
\renewcommand{\arraystretch}{0.5}
\begin{tabular}{lcccccccc}
\toprule
\textbf{Primary discovery source} & \textbf{2021} & \textbf{2022} & \textbf{2023}
& \textbf{2024} & \textbf{2025} & \textbf{2026$^{a}$} & \textbf{Total} & \textbf{\%} \\
\midrule
IEEE Xplore & 1 & 0 & 4 & 4 & 6 & 5 & \textbf{20} & \textbf{6.9} \\
ACM Digital Library & 0 & 1 & 0 & 0 & 2 & 0 & \textbf{3} & \textbf{1.0} \\
ScienceDirect & 0 & 0 & 0 & 2 & 0 & 0 & \textbf{2} & \textbf{0.7} \\
MDPI & 1 & 3 & 6 & 8 & 5 & 3 & \textbf{26} & \textbf{9.0} \\
SpringerLink & 0 & 2 & 1 & 4 & 3 & 2 & \textbf{12} & \textbf{4.2} \\
Scopus & 19 & 32 & 40 & 49 & 50 & 13 & \textbf{203} & \textbf{70.2} \\
Other methods & 0 & 3 & 4 & 8 & 7 & 1 & \textbf{23} & \textbf{8.0} \\
\midrule
\textbf{Total} & \textbf{21} & \textbf{41} & \textbf{55} & \textbf{75} & \textbf{73}
& \textbf{24} & \textbf{289} & \textbf{100.0} \\
\textbf{\%} & \textbf{7.3} & \textbf{14.2} & \textbf{19.0} & \textbf{26.0}
& \textbf{25.3} & \textbf{8.3} & \textbf{100.0} & \\
\bottomrule
\end{tabular}
\vspace{0.15cm}
\footnotesize \textit{Note}: Database rows are mutually exclusive and represent each publication's primary discovery source rather than every database in which the publication was indexed. Publications initially identified through the earlier
five-source search retain their original non-Scopus discovery source when subsequently matched in Scopus. The other-methods pathway comprises publications added through backward-reference reconciliation after deduplication against the database-derived corpus. Consequently, the six database rows comprise 266
included publications, and the other-methods pathway contributes 23, yielding the final included corpus of \(N=289\). Percentages may not sum exactly because of rounding. \(^{a}\)The 2026 observation covers January through June and is
not directly comparable with a complete calendar year.
\end{table}

Scopus was the primary discovery pathway for 203 publications (70.2\%), the
five earlier sources contributed 63 (21.8\%), and backward-reference
reconciliation contributed 23 (8.0\%). This is a provenance distribution,
not publisher concentration, because Scopus indexes multiple publishers.
Coverage restrictions, source concentration, and retrieval-related selection
bias are discussed in Section~\ref{subsec:review_limitations}.

\subsection{Data Extraction and Five-Dimensional Coding}
\label{subsec:extraction_coding_quality}

Corpus-level metadata and D1--D5 taxonomy variables were recorded for all
\(N=289\) included publications. Experiment-level variables were extracted
for the \(n=231\) original experimental studies and covered datasets,
preprocessing, GAN configuration, comparators, metrics, validation protocols,
ablations, repetitions and seeds, uncertainty/statistics, implementation
details, reproducibility artifacts, deployment evidence, numerical results,
and evidence locations.

Applicable but unreported experiment-level information was coded as
\emph{not reported} (NR); experiment-specific fields for the other 58 included
publications were coded as \emph{not applicable}. Each publication received
one primary D1--D5 code, while secondary tags retained hybrid characteristics
without inflating primary frequencies. Evidence maturity records the strongest
demonstrated validation setting from E1 (offline proof-of-concept) to E5
(sustained pilot or live operational validation). Operational definitions and
ambiguous-case decisions are preserved in Supplementary Workbook~S1
~\cite{Ndayipfukamiye2026_GANProtocol}.

\subsection{Methodological Quality and Reporting Completeness}
\label{subsec:quality_appraisal}

Eight binary criteria were applied to the \(n=231\) original experimental
studies: cybersecurity objective (Q1), GAN and defensive role (Q2), dataset
and protocol information (Q3), comparator or ablation (Q4), claim-appropriate
metrics with uncertainty/statistical support (Q5), implementation and
hyperparameter detail (Q6), study-specific reproducibility artifacts (Q7),
and limitations or threats to validity (Q8). A criterion scored 1 only when
explicitly supported by the full text. High, moderate, and low reporting
quality were defined as \(6\!-\!8/8\) (\(\geq75\%\)),
\(4\!-\!5/8\) (50--74\%), and \(0\!-\!3/8\) (\(<50\%\)),
respectively; quality scores were not eligibility criteria.

Among the \(n=231\) experimental studies, 109 (47.2\%) were high quality,
102 (44.2\%) moderate, and 20 (8.7\%) low. The remaining 58 publications
were marked not applicable for experiment-level appraisal rather than assigned
zero scores.

\subsection{Quantitative and Thematic Synthesis}
\label{subsec:synthesis}

RQ1 uses descriptive corpus and experiment-level indicators. RQ2 examines the
D1--D5 distributions and three prespecified cross-tabulations:
D1$\times$D2, D3$\times$D5, and D1$\times$D4. Pearson \(\chi^2\),
descriptive and bias-corrected Cram\'er's \(V\), adjusted residuals, and
fixed-margin permutation tests are used as corpus-level association diagnostics.
Permutation results are not interpreted as population-level inference beyond
the publications identified by this review.

RQ3 follows an outcome-specific SWiM strategy within the \(n=231\)
experimental studies. Outcomes with different substantive meanings are not
collapsed into a universal effect. Of 150 candidate comparisons sharing
nominal dataset and metric labels, none satisfied both split and preprocessing
equivalence requirements, and six of the 150 were invalid self-contrasts.
Accordingly, no pooled estimate, aggregate effect direction, or vote count is
reported. RQ4 synthesizes validity, attribution, reproducibility, and
deployment barriers, while RQ5 translates these findings into research and
engineering priorities.

\subsection{Data Management, Reproducibility, and Protocol Status}
\label{subsec:data_management}

Supplementary Workbook~S1 preserves the audit trail from raw bibliographic
inputs through screening, eligibility, D1--D5 coding, Q1--Q8 appraisal,
experiment-level extraction, matched-contrast auditing, figure data, and
validation checks~\cite{Ndayipfukamiye2026_GANProtocol}. The protocol was
not prospectively registered. The OSF record instead preserves a retrospective
protocol snapshot and revision history, including the Scopus expansion and
SpringerLink query-validation amendment; it is not presented as
preregistration. This limitation is considered in
Section~\ref{subsec:review_limitations}.

\subsection{Use of Computational and AI-Assisted Tools}
\label{subsec:AI-use}

Language-model-assisted tools were used only for non-decisional metadata
normalization, duplicate flagging, evidence-location support, and registry
consistency checks. No automated tool made or altered an eligibility,
screening, taxonomy-coding, or quality-appraisal decision; all assisted
outputs were reviewed and could be overridden by the primary reviewer or
verifier. Any AI assistance in manuscript drafting is disclosed separately
in accordance with journal policy.


\section{Five-Dimensional Taxonomy and Integrated Design Space}
\label{sec:taxonomy}

This section organizes the reviewed evidence into a five-dimensional design space linking defensive function, GAN architecture, cybersecurity domain, threat or data condition, and evidence and deployment context.

\subsection{Taxonomy Purpose and Coding Logic}
\label{subsec:taxonomy_purpose}

The taxonomy operationalizes RQ2 by representing GAN-based cybersecurity
defense as a multidimensional design space rather than a list of architectures
or applications. Whereas prior reviews have largely treated GAN architectures,
intrusion-detection applications, synthetic cyberattack data, and anomaly
detection separately~\cite{YinkaBanjo2020,10187144,electronics13020322,
ALAJLAN20242053,LIM2024103733,Alauthman2026,SAEED2026128978,
10.1145/3453158,BOUNTAKAS2023100573,10005100}, the present framework
integrates these perspectives across five dimensions.

The unit of analysis is each of the \(N=289\) included full-text publications.
Each receives one primary code for defensive function (D1), GAN architecture
(D2), cybersecurity domain (D3), threat or data condition (D4), and strongest
demonstrated evidence maturity (D5), respectively asking: \emph{What role does
the GAN perform? How is it realized? Where is it applied? Which security risk
or data limitation is addressed? Under what conditions is the contribution
validated?} Publications outside the review eligibility criteria do not enter
taxonomy frequencies.

\begin{figure}[htbp]
    \centering
    \includegraphics[width=\linewidth]{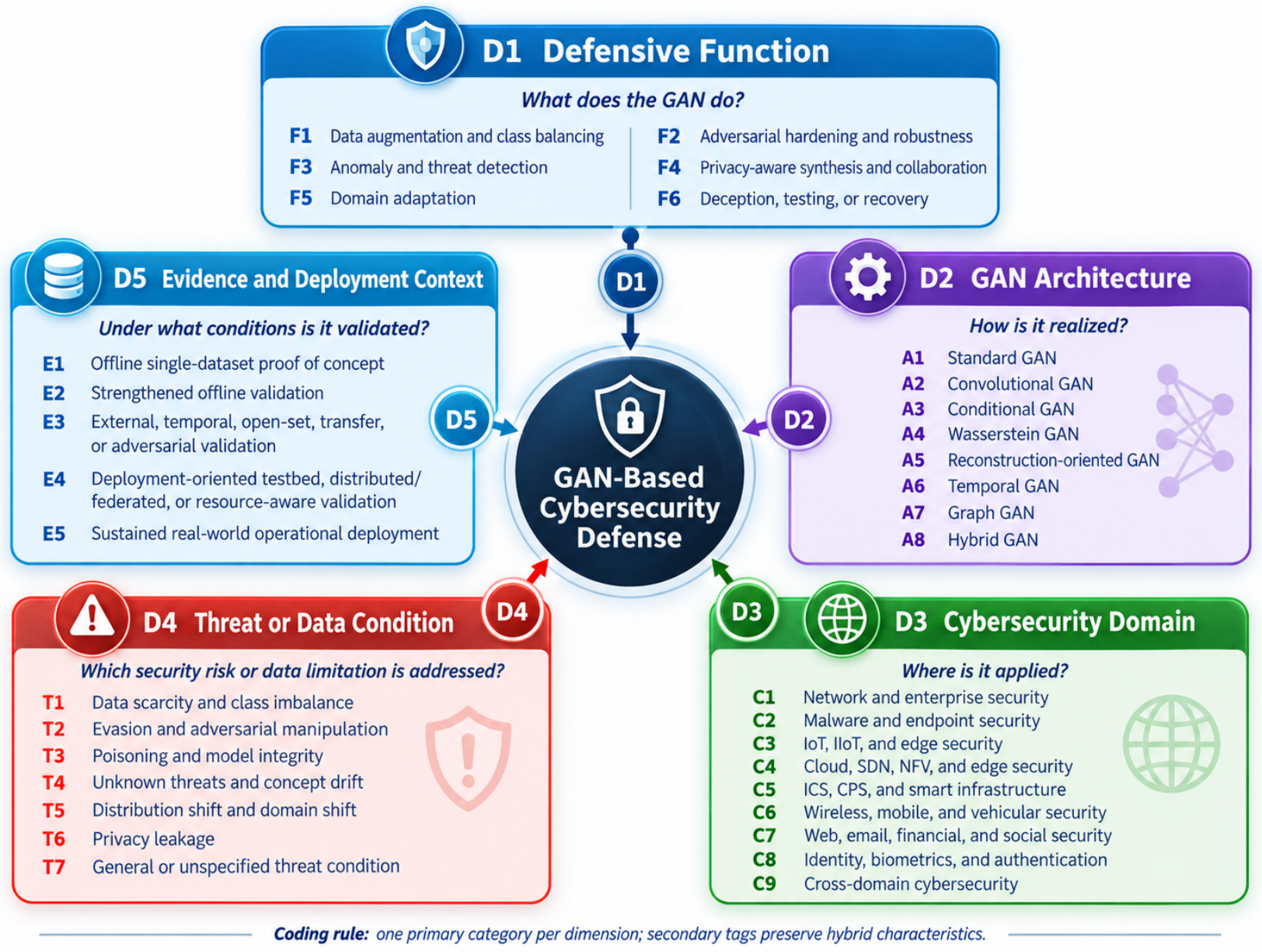}
    \caption{\textbf{Five-dimensional taxonomy of GAN-based cybersecurity defense.} Each included publication is mapped by defensive function, primary GAN architecture, cybersecurity domain, threat or data condition, and strongest demonstrated evidence maturity. One primary category is assigned per dimension, while secondary tags preserve hybrid characteristics.}
    \label{fig:five_dimensional_taxonomy}
\end{figure}

One primary category is assigned per dimension, while secondary tags preserve
hybrid functions, architectural modifiers, additional threat attributes, and
multi-domain settings. Primary coding follows the publication's principal
cybersecurity contribution rather than the component producing the highest
metric value. Thus, a GAN used principally to generate minority-class samples
for a downstream detector is coded as augmentation (F1), whereas a GAN whose
reconstruction error provides the anomaly signal is coded as detection (F3).
The coding decision rules, including the adjudication of ambiguous cases,
are documented in Supplementary Workbook~S1, together with the operational
definitions and validation checks used to maintain consistency across the
D1--D5 taxonomy. Figure~\ref{fig:five_dimensional_taxonomy} summarizes the framework.

Architecture coding draws on established standard, convolutional, conditional,
Wasserstein, reconstruction-oriented, temporal, graph, and hybrid GAN families
~\cite{NIPS2014_f033ed80,radford2015unsupervised,
DBLP:journals/corr/MirzaO14,pmlr-v70-arjovsky17a,NIPS2017_892c3b1c,
SCHLEGL201930,NEURIPS2019_c9efe5f2}, while secondary tags retain features such
as conditioning, gradient penalties, attention, autoencoders, transformers,
graph components, federated learning, differential privacy, and explainability
~\cite{ZOUHRI2025107882,HAMOUDA2024101149,Khalil2026,11555542}.
Foundational GAN papers define coding mechanisms but are not counted in the
cybersecurity corpus.

Coding follows an \emph{evidence-over-claim} principle: labels such as
``real-time,'' ``lightweight,'' ``privacy-preserving,'' or ``operational'' are
accepted only when supported by reported evidence. Metrics such as accuracy,
F1, AUC, latency, energy, and communication cost are treated as evidence
variables rather than taxonomy categories. Complete full texts informed D1--D5
coding for all \(N=289\) publications, while experiment-specific variables were
extracted separately for the \(n=231\) original experimental studies. Detailed
definitions, secondary tags, and publication-level coding decisions are
preserved in Supplementary Workbook~S1
~\cite{Ndayipfukamiye2026_GANProtocol}.


\subsection{Taxonomy Overview and Category Structure}
\label{subsec:taxonomy_overview}

As summarized in Figure~\ref{fig:five_dimensional_taxonomy}, the taxonomy maps
each included publication across five complementary dimensions: defensive
function, GAN architecture, cybersecurity domain, threat or data condition,
and evidence and deployment context. Together, these dimensions connect
technical design with the security problem, protected environment, and
strength of supporting evidence.

Table~\ref{tab:taxonomy_overview} summarizes the primary categories in each dimension. Detailed operational definitions and secondary tags are provided in Supplementary Workbook~S1~\cite{Ndayipfukamiye2026_GANProtocol}, while
Sections~\ref{subsec:dim1_function}--\ref{subsec:dim5_evidence} examine the five dimensions in detail.

\begin{table*}[!ht]
\centering
\caption{Primary category structure of the five-dimensional taxonomy.}
\label{tab:taxonomy_overview}
\scriptsize
\setlength{\tabcolsep}{1.5pt}
\renewcommand{\arraystretch}{0.8}

\begin{tabular}{p{2.0cm}p{2.5cm}p{10.0cm}p{2.5cm}}
\toprule
\textbf{Dimension} &
\textbf{Analytical question} &
\textbf{Primary categories} &
\textbf{Key References} \\
\midrule

\textbf{D1: Defensive function}
&
What does the GAN do?
&
\textbf{F1} Data augmentation and class balancing;
\textbf{F2} adversarial hardening and robustness;
\textbf{F3} anomaly and threat detection;
\textbf{F4} privacy-aware synthesis and collaboration;
\textbf{F5} domain adaptation;
\textbf{F6} deception, testing, or recovery
&
\cite{ANDRESINI2021108,QIN2024110655,MISHRA2024493,TABASSUM2022299,Yang2025CEGAN}
\\

\addlinespace

\textbf{D2: GAN architecture}
&
How is it realized?
&
\textbf{A1} Standard GAN;
\textbf{A2} convolutional GAN;
\textbf{A3} conditional GAN;
\textbf{A4} Wasserstein GAN;
\textbf{A5} reconstruction-oriented GAN;
\textbf{A6} temporal GAN;
\textbf{A7} graph GAN;
\textbf{A8} hybrid GAN
&
\cite{NIPS2014_f033ed80,radford2015unsupervised,
DBLP:journals/corr/MirzaO14,pmlr-v70-arjovsky17a,SCHLEGL201930}
\\

\addlinespace

\textbf{D3: Cybersecurity domain}
&
Where is it applied?
&
\textbf{C1} Network and enterprise security;
\textbf{C2} malware and endpoint security;
\textbf{C3} IoT, IIoT, and edge security;
\textbf{C4} cloud, SDN, NFV, and edge security;
\textbf{C5} ICS, CPS, and smart infrastructure;
\textbf{C6} wireless, mobile, and vehicular security;
\textbf{C7} web, email, financial, and social security;
\textbf{C8} identity, biometrics, and authentication;
\textbf{C9} cross-domain cybersecurity
&
\cite{Lei2025FogGAN,AlEssa2026PEAdversarialGAN,
Lilhore2026VehicularGAN,Girish2026GraphGANFraud}
\\
\addlinespace
\textbf{D4: Threat or data condition}
&
Which security risk or data limitation is addressed?
&
\textbf{T1} Data scarcity and class imbalance;
\textbf{T2} evasion and adversarial manipulation;
\textbf{T3} poisoning and model integrity;
\textbf{T4} unknown threats and concept drift;
\textbf{T5} distribution shift and domain shift;
\textbf{T6} privacy leakage;
\textbf{T7} general or unspecified threat condition
&
\cite{10.1145/3453158,ZHU2022485,AlEssa2026PEAdversarialGAN,
Priya2026SpectreCGAN}
\\
\addlinespace
\textbf{D5: Evidence and deployment context}
&
Under what conditions is it validated?
&
\textbf{E1} Offline single-dataset proof of concept;
\textbf{E2} strengthened offline validation;
\textbf{E3} external, temporal, open-set, transfer, or adversarial validation;
\textbf{E4} deployment-oriented testbed, distributed/federated, or
resource-aware validation;
\textbf{E5} sustained real-world operational deployment
&
\cite{s26030757,Devi2024LightweightDCGAN,Durga2026TinyGAN,
AlSarray2025GANCLSTO}
\\

\bottomrule
\end{tabular}

\footnotesize
\textit{Note:} Each included publication receives one primary category per dimension; secondary tags preserve hybrid mechanisms, additional functions, threat attributes, and supplementary evaluation settings. Representative references may include contextual foundational sources, which do not enter corpus frequencies.
\end{table*}

\begin{figure}[!ht]
\centering
\includegraphics[width=\linewidth]
{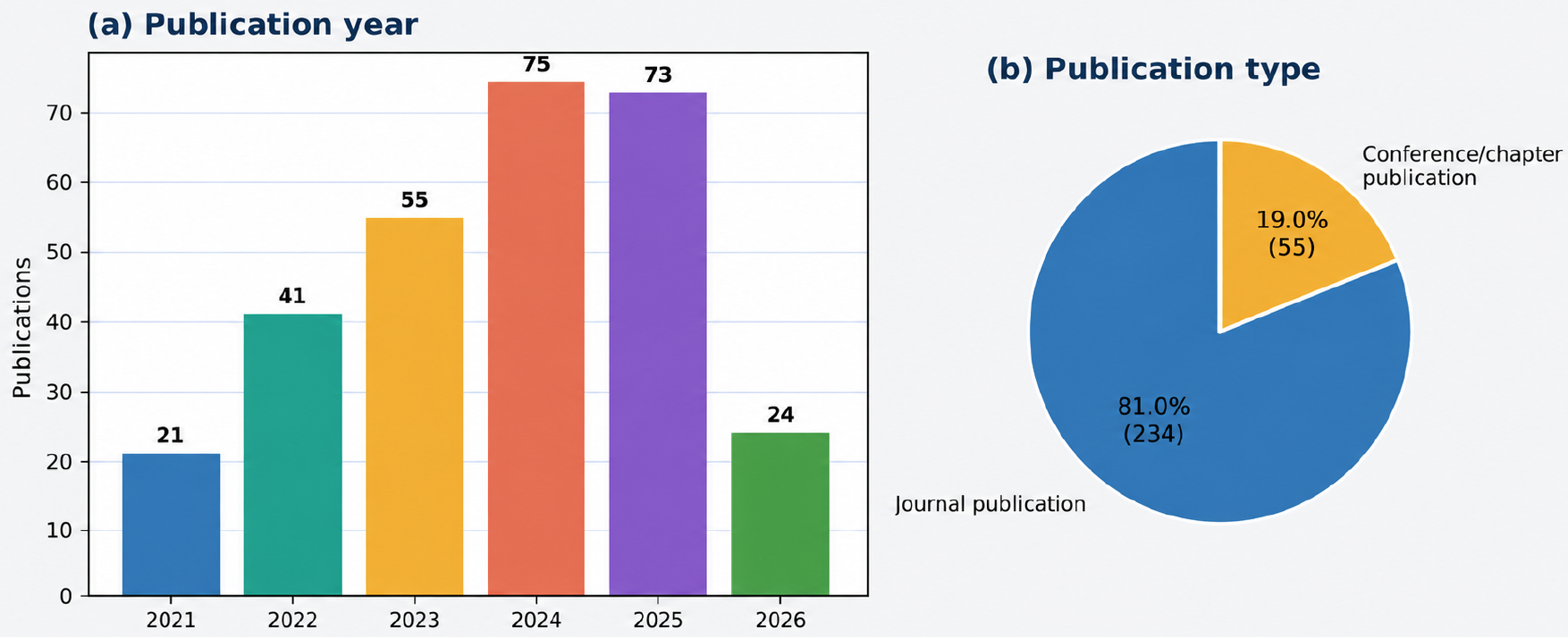}
\caption{\textbf{Temporal and publication-type distribution of the included corpus (\(N=289\)).} Annual counts cover January 2021--June 2026 and distinguish journal publications from conference/chapter publications.}
\label{fig:temporal_distribution}
\end{figure}

Figure~\ref{fig:temporal_distribution} shows sustained publication growth, from 21 publications in 2021 to 75 in 2024 and 73 in 2025; 24 publications were identified through June~2026. Overall, 172 publications (59.5\%) appeared from 2024 onward. Journal publications accounted for 234 (81.0\%), compared with 55 conference/chapter publications (19.0\%). Because the 2026 stratum covers January--June only, it is not directly comparable with complete calendar years; this constraint is considered in Section~\ref{subsec:review_limitations}.

\subsection{Dimension 1: Defensive Function--What Does the GAN Do?}
\label{subsec:dim1_function}

\begin{figure}[!ht]
\centering
\includegraphics[width=\linewidth]{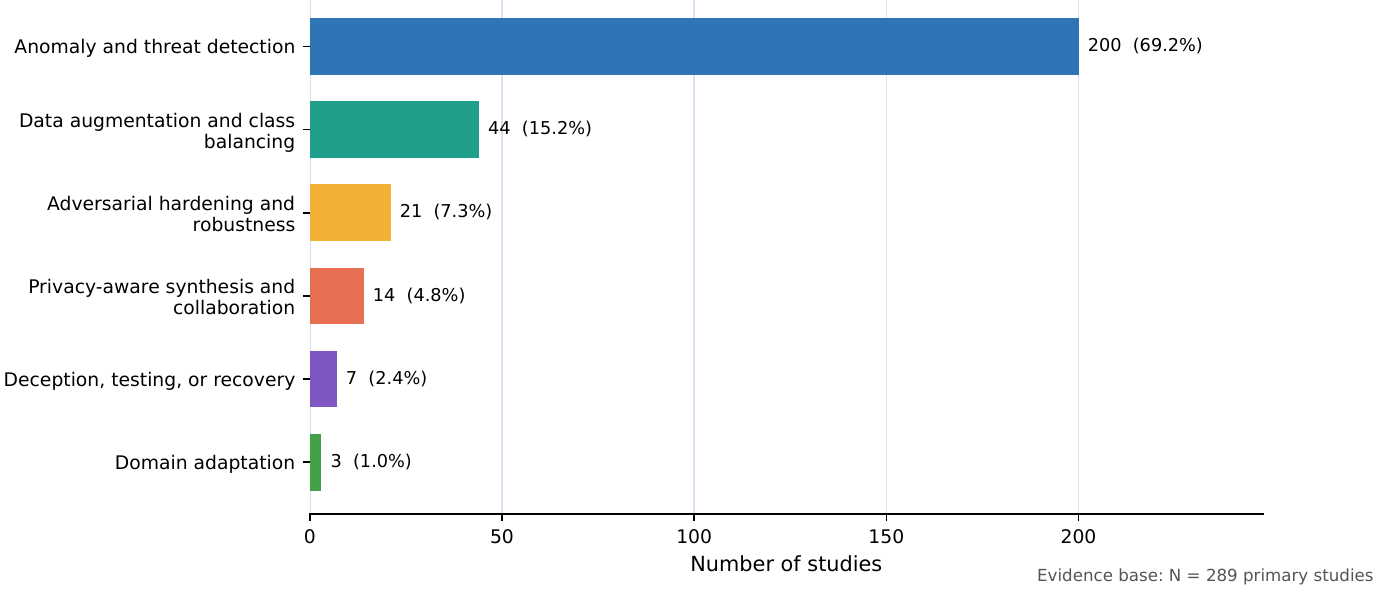}
\caption{\textbf{Primary defensive-function distribution (\(N=289\)).} Each included publication contributes one primary D1 code; bar labels report publication counts.}
\label{fig:dim1_function}
\end{figure}

D1 identifies the GAN's primary defensive role and most directly operationalizes
the review's distinction between \emph{threat detection} and \emph{mitigation}.
Figure~\ref{fig:dim1_function} shows a pronounced concentration in anomaly and
threat detection (F3: 200 publications, 69.2\%). Mitigation-oriented functions
comprise adversarial hardening and robustness (F2: 21, 7.3\%), privacy-aware
synthesis and collaboration (F4: 14, 4.8\%), domain adaptation (F5: 3, 1.0\%),
and deception, testing, or recovery (F6: 7, 2.4\%), together accounting for
45 publications (15.6\%). Data augmentation and class balancing (F1: 44,
15.2\%) is retained as a distinct enabling function because its immediate role
is to improve defensive training data rather than directly detect or mitigate
a threat.

These are corpus-level prevalence counts, not estimates of comparative
effectiveness. The distinction is important because each function addresses a
different defensive failure and therefore requires different evidence for a
credible security claim.

\subsubsection{F1: Data augmentation and class balancing}
\label{subsubsec:D1F1}

GAN augmentation generates additional training observations, particularly for
rare attack classes, to improve a downstream defensive model. It is the largest
non-detection enabling function in the corpus and spans network intrusion,
power-system monitoring, malware, fraud, IoT, and hardware-attack detection
~\cite{ANDRESINI2021108,SRIVASTAVA2023103432,ZHAO2024104005,
AGARWAL2024115603,Islam2026PAACGAN,Zhu2026BADGAN}. Conditional,
Wasserstein, graph-based, and application-specific approaches extend this role
to imbalanced NIDS, banking transactions, Spectre attacks, and advanced
metering infrastructure
~\cite{Yang2025CEGAN,Yao2025CWGANGP,Girish2026GraphGANFraud,
Priya2026SpectreCGAN,electronics15010235}.

Across the corpus, augmentation is implemented through oversampling, synthetic
attack generation, and imbalance-aware training in network, IoT, malware, and
financial settings
~\cite{ShouniBarkah2023Data,Iftikhar2025Enhanced,Fukas2022Augmenting,
Langevin2022Generative,Xie2023Credit,Ding2023Credit,Zhao2023CTCN,
Ghaleb2023Ensemble,Zhao2024Advancing,Gudivaka2024Improved,
10873111,Ma2025Addressing,Alrasheedi2025Advanced,
Zhang2025Credit,10956249}. Generative realism alone is not treated
as defensive effectiveness: the relevant claim is whether synthetic data
improve downstream performance on untouched real observations without leakage,
duplication, or distortion of class structure. Experiment-level evidence for
this claim is examined in Section~\ref{sec:empirical_synthesis}.

\subsubsection{F2: Adversarial hardening and robustness}
\label{subsubsec:F2D1}

In adversarial hardening, the GAN generates evasive, boundary-region, or
otherwise difficult observations that expose model weaknesses and support
retraining, stress testing, or recovery. Applications include botnet-detector
hardening, keystroke-injection defense, adversarial malware classification,
and post-attack recovery
~\cite{9439890,10.1145/3700838.3700871,AlEssa2026PEAdversarialGAN,
11075744,Feizi2024}.

Additional publications evaluate adversarially generated attacks or
robustness-oriented retraining in malware, network, and autonomous-system
settings
~\cite{Kumaran2023Adversarial,Wang2025Improving,
Gopalakrishnan2025Addressing,Zhang2021Enhanced,Wang2021LSGAN,
Reilly2023Robustness,Martinelli2024Evaluating,Ali2025AEDGAN,
Dhingra2026Image,Badrinath2026Improved}. Classification as F2 requires an
implemented defensive response and post-defense evaluation rather than attack
generation alone; attack-only studies contribute threat-model context instead
~\cite{ZHU2022485,11063505}.

Accordingly, credible F2 evidence concerns the robustness--utility trade-off
under an explicit attacker model: performance should be evaluated both before
and after attack, with attacker knowledge, budget, transferability, and
generalization beyond the training attack made explicit. Resistance to one
fixed generator does not establish robustness to adaptive adversaries.

\subsubsection{F3: Anomaly and threat detection}
\label{subsubsec:F3D1}

F3 is the dominant defensive function in the corpus. Here the GAN participates
directly in detection through discriminator features, latent representations,
reconstruction error, adversarially learned anomaly scores, or features passed
to a downstream detector
~\cite{Albalawi2026GANIDS,Narayan2026ZIDS,Barpute2026AIDRRGAN,
Saini2026CyResWadc,Thirumalaisamy2026SmartFarmingGAN,
Wang2025GraphACGAN}. Applications span IoT, DDoS, industrial,
cyber--physical, 5G, and distributed anomaly detection
~\cite{MISHRA2024493,11416973,Bhutani2025SADGAN,11550383}, while
federated formulations extend detection to 6G, industrial-control, and
multi-domain settings~\cite{11037504,11569888,10988602}.

The broader corpus confirms that this function encompasses conventional NIDS,
hybrid deep detectors, one-class anomaly models, and specialized
network-security pipelines
~\cite{Liu2021GAN,Shi2021Cyber,Girish2021DadGAN,Kim2020Generative,
Macwan2021GIDS,Zhou2021Indoor,Das2021Network,Jeong2021Poster,
Zhang2021Research,Chao2021Research,Meddahi2021SIP,
Chen2021Variational,Fang2022Block,Kar2022Integrated,
Idrissi2022Unsupervised,Kim2022Early,Geng2022Effective,
Hallaji2022Embedding,Balaji2022HybridDeep,Alqarni2022Improving,
Nie2022Intrusion,Zou2022Intrusion,Iliyasu2022N,Zhou2022Network,
Soleymanzadeh2022Analysis,LI2022109283,Li2022Generative,
Rahman2023Quantum,Ahmad2023Boundary,Preethi2023Cycle}.
Subsequent systems combine GANs with graph learning, temporal modeling,
explainability, ensembles, and optimization
~\cite{Jha2023GAN,Hussain2023GAN,Pandey2023Generative,
BENCHAMA2023Novel,Guo2023RegraphGAN,Zhao2023Research,
Sagar2023SECURITY,DeAraujoFilho2023Unsupervised,Han2023XA,
Luo2024Cybersecurity,Schoen2024Tale,PRamya2024Advanced,Rao2024AI,
Sarika2024AI,Pawar2024Advanced,Wen2024Application,
Vallabhaneni2024Detection,Krishnasamy2024DIWGAN,Feng2024Efficient,
Yu2024Enhanced,Meenakshi2024Enhancing,Arafah2024Evaluating,
Constantin2025Exploring,Li2024FlowGANAnomaly,Nguyen2024Generative,
Veksler2024Good,Qin2024GPIDS,AlMilli2024Intrusion,Zeeshan2024Trans,
Changala2024Generative}. Recent work continues this trajectory through
explainable, optimized, lightweight, and zero-day IDS designs
~\cite{Tarek2025Keyless,Fan2025AGASI,Chinnasamy2025Explainable,
SaravanaRam2025Intrusion,Allagi2025Enhanced,Peppes2025Evaluating,
AlAjlan2025GAN,Sekhar2025Generative,Cirillo2025Intrusion,
Li2025Mitigating,Alyoubi2025Network,Albalawi2025Strategic,
Liu2025Symmetric,Kang2025VAE,Chandrakala2026Distillation,
Esfahani2026Enhancing,Adiban2021STEP,Ahsan2022Comparative,
Striuk2023Cross,Tang2025Swarm,Alotaibi2026Enhancing,
Badakhshan2026Generative}.

The central evidential question is not whether a detector achieves a high
closed-set score, but whether its GAN-derived signal separates legitimate and
malicious behavior under conditions relevant to the claimed threat. Zero-day
or unknown-threat claims therefore require attack-family, temporal, device, or
external-data separation rather than random closed-set partitions. Threshold
selection, contamination of normal data, false-positive burden, and comparison
with strong non-GAN detectors remain critical to interpreting F3 evidence.

\subsubsection{F4: Privacy-aware synthesis and collaboration}
\label{subsubsec:F4D1}

F4 uses GANs to generate shareable synthetic records, support collaborative
learning without centralizing raw observations, or incorporate explicit
privacy mechanisms. Approaches combine GANs with federated learning,
differential privacy, model-inversion defense, and collaborative intrusion
detection~\cite{TABASSUM2022299,Feizi2024,10184476,10437148,Ullah2026}.
More recent systems examine federated GAN augmentation in fog, IoT, and
resource-constrained IoMT environments~\cite{Lei2025FogGAN,Nguyen2025FederatedWGAN,Durga2026TinyGAN}.

The evidence also covers federated generation, privacy-aware augmentation, and
distributed or incremental protection
~\cite{Quyen2022Federated,Devi2022Intrusion,Yang2022Lightweight,
Wang2022Federated,Qiao2024Privacy,Jiang2024OFPP,Yao2024Privacy,
Murmu2024Reliable,Li2025CryptoGAN,Masse2025Deception,
Islam2025Exposing,Iantorno2025Fuzzy}. Importantly, federation or synthetic-data
generation is not itself treated as proof of privacy. A defensible F4 claim
requires an explicit privacy mechanism or leakage threat together with
measurement of privacy and defensive utility; distributed settings additionally
raise client-heterogeneity, communication-cost, and malicious-client concerns.
The relevant evidence is therefore a privacy--utility--communication trade-off,
not data locality alone.

\subsubsection{F5: Domain adaptation}
\label{subsubsec:F5D1}

Domain-adaptation approaches use adversarial feature alignment or target-like
generation to reduce degradation between source and deployment environments.
Representative applications include few-shot IoT intrusion detection,
cross-domain malware detection, and IoT--cloud security
~\cite{MA2023719,BHARDWAJ2024103769,NAZIM2025101773}.

F5 is distinct from ordinary augmentation because a source--target discrepancy
is explicit in both the method and the evaluation. Evidence therefore depends
on meaningful separation between domains, for example, networks,
organizations, devices, periods, or attack families, and comparison against
source-only and non-GAN adaptation baselines. A random split of one dataset
does not establish domain adaptation.

\subsubsection{F6: Deception, testing, or recovery}
\label{subsubsec:F6D1}

F6 uses generated decoys, probes, simulated attacks, or recovery-oriented
observations to alter, evaluate, or restore attacker--defender interaction.
Applications include reconnaissance deception in industrial control systems,
generative IoT honeypots, autonomous-driving anomaly recovery, and
post-adversarial restoration of 5G intrusion detectors
~\cite{QIN2024110655,11245604,10430114,11075744,
Kumar2026CyberDeception}. Generating attack traffic alone is insufficient for F6 classification. The generated artifact must support an implemented defensive objective such as attacker misdirection, vulnerability testing, improved detection, containment, or recovery. Appropriate outcomes therefore extend beyond classification accuracy to engagement time, deception success, vulnerability coverage, time-to-detection, recovery time, availability, and defender cost.

Taken together, D1 reveals a strongly detection-oriented evidence base.
Threat detection accounts for 200/289 publications (69.2\%), compared with
45/289 (15.6\%) addressing mitigation-oriented functions and 44/289 (15.2\%)
using augmentation as an enabling data-level defense. These proportions
indicate the concentration of research activity rather than comparative
effectiveness across defensive functions.


\subsection{Dimension 2: GAN Architecture--How Is the Defense Realized?}
\label{subsec:dim2_architecture}

\begin{figure}[!ht]
\centering
\includegraphics[width=\linewidth]{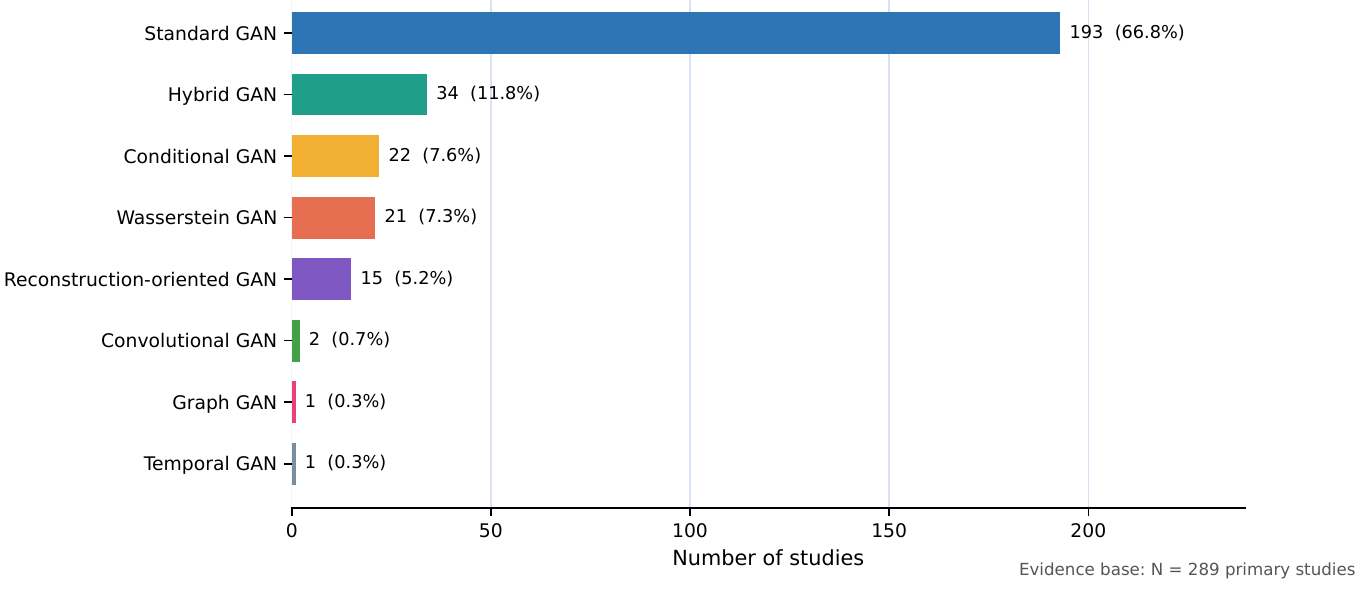}
\caption{\textbf{Primary GAN-architecture distribution (\(N=289\)).}
Each included publication contributes one primary D2 code; architectural modifiers are retained as secondary tags.}
\label{fig:dim2_architecture}
\end{figure}

D2 identifies the primary adversarial formulation used to realize the defense, independently of defensive function. Figure~\ref{fig:dim2_architecture} shows a
strong concentration in standard GANs (193 publications, 66.8\%), followed by hybrid GANs (34, 11.8\%), conditional GANs (22, 7.6\%), Wasserstein GANs (21, 7.3\%), and reconstruction-oriented GANs (15, 5.2\%). Convolutional, temporal, and graph GANs are uncommon as primary architectures.

The corpus also contains robustness-oriented, composition-aware, quantum, graph, CycleGAN, and other specialized variants
~\cite{Yang2023Robust,Zheng2023Composition,AlKateb2024CryptoGenSec,
Singha2025Zero,Yadkar2026CycleGAN,Rahman2023Quantum,
Guo2023RegraphGAN,Kang2025VAE}. These architectural labels describe how the defense is implemented, not how effective it is; interpretation therefore depends on the defensive function, data structure, comparator, and evaluation setting in which the architecture was tested.

\subsubsection{A1--A2: Standard and convolutional GANs}
\label{subsubsec:A1A2D2}

Standard GANs retain the basic generator--discriminator objective and support
general augmentation, feature learning, and proof-of-concept defenses. They
also provide a useful reference for determining whether conditioning,
stability-oriented objectives, or additional components add measurable value
~\cite{NIPS2014_f033ed80,ANDRESINI2021108}.

Convolutional GANs introduce spatial inductive biases and are most appropriate
when traffic, executable bytes, sensor signals, or biometric observations are
represented as images or structured matrices
~\cite{radford2015unsupervised,computers13060154,
Devi2024LightweightDCGAN}. Their validity depends on whether spatial proximity
preserves meaningful security structure; visual realism alone does not establish
protocol validity, malware functionality, or downstream defensive utility.
A publication is coded as A1 or A2 only when no more specific adversarial
formulation governs the method.

\subsubsection{A3: Conditional GANs}
\label{subsubsec:A3D2}

Conditional GANs incorporate labels or contextual variables into generation,
enabling targeted synthesis of attack families, device states, traffic classes,
or other security conditions~\cite{DBLP:journals/corr/MirzaO14}. They are
widely used for minority-class augmentation and controlled intrusion-data
generation
~\cite{app13042576,s23125644,math14081264,Yang2025CEGAN,
Priya2026SpectreCGAN}, including protocol-aware augmentation and graph-enhanced
ACGAN designs~\cite{Islam2026PAACGAN,Wang2025GraphACGAN}.

Conditioning improves controllability but does not guarantee fidelity or
diversity. Its defensive value depends on label quality, within-class
heterogeneity, real sample availability, and downstream performance on genuine
test observations. Specialized tabular normalization and sampling mechanisms
are retained as secondary architectural tags.

\subsubsection{A4: Wasserstein GANs}
\label{subsubsec:A4D2}

Wasserstein GANs replace the original adversarial divergence with a critic-based
objective intended to provide more informative gradients, while WGAN-GP adds a
gradient penalty to improve enforcement of the Lipschitz constraint
~\cite{pmlr-v70-arjovsky17a,NIPS2017_892c3b1c}. They appear in imbalanced
intrusion detection, federated augmentation, industrial anomaly detection, and
security-data synthesis
~\cite{SRIVASTAVA2023103432,electronics14204068,Yao2025CWGANGP,
Nguyen2025FederatedWGAN,Singh2025SideChannelWGAN}.

Their principal architectural contribution is training stability rather than
automatic defensive superiority. Consequently, convergence or smoother losses
are relevant only when they translate into improved security utility,
diversity, robustness, or efficiency under appropriate baselines.

\subsubsection{A5: Reconstruction-oriented GANs}
\label{subsubsec:A5D2}

Reconstruction-oriented GANs combine adversarial learning with autoencoding,
latent inversion, feature matching, or reconstruction losses. Anomaly scores
may therefore derive from reconstruction error, latent inconsistency,
discriminator responses, or combinations of these signals
~\cite{SCHLEGL201930,10.1007/978-3-030-20893-6_39}.

Applications include MQTT and IoT intrusion detection, 5G-core monitoring,
cyber--physical anomaly detection, and autonomous-driving recovery
~\cite{BOPPANA2023105805,MISHRA2024493,11550383,
Bhutani2025SADGAN,10430114}. Their effectiveness depends particularly on
normal-data purity, model capacity, anomaly-score construction, and threshold
selection: overly expressive models may reconstruct attacks, whereas weak
models may falsely flag legitimate variation. Open-set or temporally separated
evaluation is therefore especially important for this family.

\subsubsection{A6--A7: Temporal and graph GANs}
\label{subsubsec:A6A7D2}

Temporal GANs model ordered behavior and sequential dependencies and are
appropriate when event order is integral to traffic sequences, authentication
traces, industrial signals, malware API calls, or multistage attacks
~\cite{NEURIPS2019_c9efe5f2,11550383,10430114}. Their evaluation should
preserve chronology; the presence of timestamps alone does not justify a
temporal architecture.

Graph GANs represent nodes, edges, neighborhoods, or relational structures and
are relevant when security behavior depends on communication, transaction,
social, or asset relationships
~\cite{Wang_Wang_Wang_Zhao_Zhang_Zhang_Xie_Guo_2018,
Girish2026GraphGANFraud,Lilhore2026VehicularGAN}. Structural fidelity,
downstream graph utility, changing relationships, and scalability are therefore
central evaluation considerations. A GAN followed by a graph classifier is not
coded as A7 unless adversarial graph generation or relational modeling is
architecturally central.

\subsubsection{A8: Hybrid GANs}
\label{subsubsec:A8D2}

Hybrid GANs combine adversarial generation with autoencoders, CNNs, recurrent
models, transformers, graph networks, federated learning, differential privacy,
explainability, blockchain coordination, or optimization. Examples include
interpretable IoT detection, GAN--transformer phishing defense, federated
augmentation, lightweight IoMT security, and explainable federated IDS
~\cite{ZOUHRI2025107882,MELISARAHEBI2025103745,Lei2025FogGAN,
Durga2026TinyGAN,11250986,11007173}.

Hybridization can extend capability but complicates attribution because observed
gains may arise from the GAN, downstream classifier, preprocessing, optimization,
or increased model capacity. The A8 primary code is therefore assigned only
when the integration itself constitutes the principal architectural contribution;
otherwise, the core GAN family receives the primary code and additional
components are retained as secondary modifiers.

Taken together, D2 shows that GAN-based cybersecurity defense is realized
predominantly through standard adversarial formulations, with specialized
architectures introducing control, stability, reconstruction, sequential or
relational structure, or component integration when required by the defensive
problem. Architectural prevalence is therefore descriptive rather than
evidence of suitability or superiority; its relationship with defensive
function is examined directly in Section~\ref{subsec:cross_dimensional_analysis}.

\subsection{Dimension 3: Cybersecurity Application Domain--Where Is the Defense Applied?}
\label{subsec:dim3_domain}

\begin{figure}[!ht]
\centering
\includegraphics[width=\linewidth]{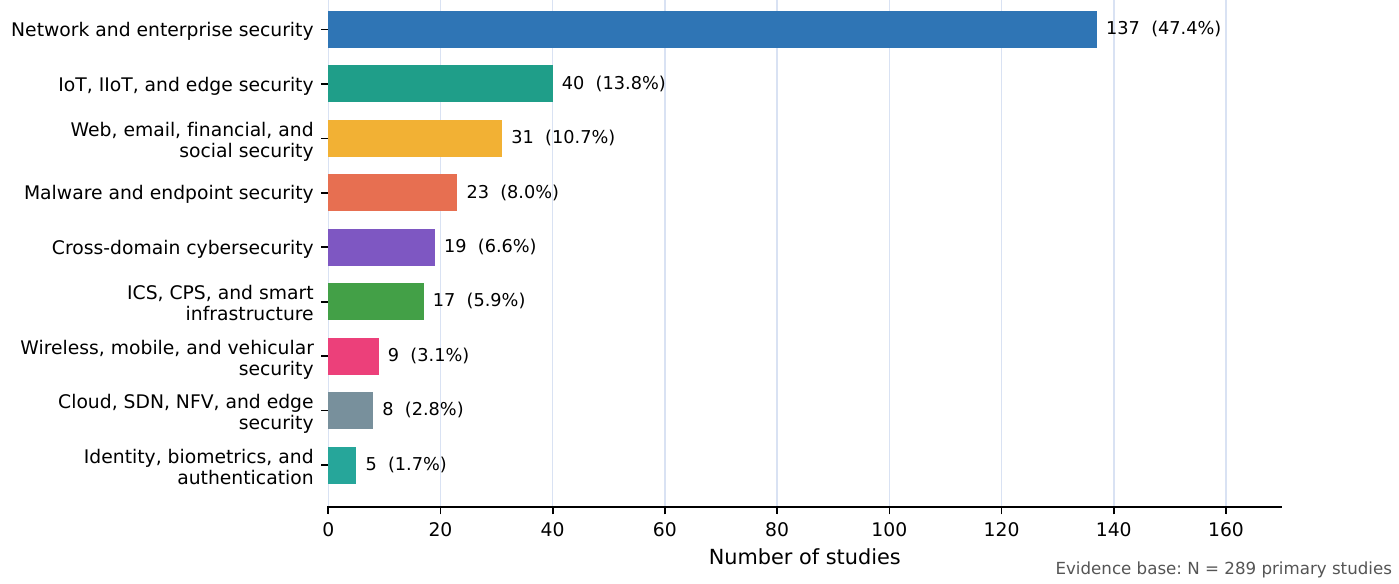}
\caption{\textbf{Primary cybersecurity-domain distribution (\(N=289\)).} Each included publication contributes one primary D3 code representing its principal protected environment.}
\label{fig:dim3_domain}
\end{figure}

D3 identifies the principal protected environment in which the GAN-based
defense is evaluated. Figure~\ref{fig:dim3_domain} shows substantial domain
concentration: network and enterprise security accounts for 137 publications
(47.4\%), followed by IoT, IIoT, and edge security (40, 13.8\%) and web,
email, financial, and social security (31, 10.7\%). Together, these three
domains represent 208/289 publications (72.0\%), indicating that the evidence
base remains concentrated in traffic-, device-, and transaction-centered
settings despite application across nine cybersecurity domains.

\subsubsection{Network and Enterprise Security}
\label{subsubsec:domain_network}

Network and enterprise security is the dominant application domain. GANs
support minority-class augmentation, traffic anomaly detection, botnet and
DDoS identification, and adversarial hardening
~\cite{ANDRESINI2021108,SRIVASTAVA2023103432,ZHAO2024104005,
electronics14204068}, with conditional and Wasserstein formulations extending
class-controlled and stability-oriented synthesis
~\cite{Yang2025CEGAN,Yao2025CWGANGP,Albalawi2026GANIDS,
Islam2026PAACGAN,Wang2025GraphACGAN}.

The principal validity boundary is generalization across flows, sessions,
hosts, campaigns, networks, and collection periods. Record-level random
splitting can place closely related observations in development and test sets;
stronger evidence therefore uses leakage-resistant partitions, real held-out
traffic, and temporal or cross-network evaluation. The numerical dominance of
this domain should consequently not be interpreted as equivalent deployment
maturity.

\subsubsection{Malware and Endpoint Security}
\label{subsubsec:domain_malware}

GANs in malware and endpoint security support malware-family augmentation,
behavioral synthesis, unknown-family detection, reconstruction-based anomaly
scoring, and adversarial hardening
~\cite{SHARMA2024122678,computers13060154,10903749,11478946,
AlEssa2026PEAdversarialGAN}. The wider evidence includes image- and
behavior-based classification, family augmentation, anomaly detection, and
adversarial training
~\cite{Zhang2021Enhanced,Wang2021LSGAN,Geetha2022Effective,
Alzahem2022Optimizing,Nguyen2023Generative,Reilly2023Robustness,
Jeyalakshmi2024Enhancing,Martinelli2024Evaluating,Carter2024IoT,
Huang2024Malware,Ali2025AEDGAN,Martin2025Deep,Wang2025ImprovingCyber,
Dhingra2026Image,Badrinath2026Improved}.

A central distinction is between classifier-oriented representations and
functionally valid malware artifacts: visual or statistical realism does not
establish executability or preservation of malicious behavior. Family,
campaign, packing, and temporal leakage also constrain zero-day claims, making
family-disjoint or temporally separated testing particularly important.

\subsubsection{IoT, IIoT, and Edge Security}
\label{subsubsec:domain_iot}

IoT, IIoT, IoMT, and edge environments combine heterogeneous devices,
resource constraints, non-IID data, and communication limitations. These
conditions motivate GAN-based augmentation, anomaly detection, privacy-aware
collaboration, deception, and federated learning
~\cite{MISHRA2024493,ZOUHRI2025107882,Ullah2026,11245604,
Thirumalaisamy2026SmartFarmingGAN}. Recent systems include fog-based
generation, federated Wasserstein augmentation, lightweight DCGAN detection,
and resource-constrained IoMT defense
~\cite{Lei2025FogGAN,Nguyen2025FederatedWGAN,
Devi2024LightweightDCGAN,Durga2026TinyGAN,Kumar2026GAIBIoD,
Kumar2026GANDQNIoT}.

The broader evidence applies conditional generation, anomaly detection,
augmentation, and hybrid learning across IoT, fog, SDN, industrial, and
vehicular settings
~\cite{Neshenko2021Behavioral,Liu2021CGAN,Sha2022Adaptive,
Huang2022Attack,Rathinavel2022Efficient,Balaji2022Hybrid,
Nukavarapu2022IKnight,Tu2022Intrusion,Ezeh2023SDN,
WeiChin2023Anomaly,Wang2023Classification,Chu2023Improving,
Poongodi2023Intrusion,Badawy20246G,Balaji2024GAN,
Kumaran2024Intrusion,10960662,
Almasre2024Create,Desanamukula2024Enhancing,Mbow2024Evading,
Alhoraibi2024Generative,Wang2024Intrusion,Sun2024MTS,
Yang2024WSN,Ahmad2025GAN,Zheng2025IoT,Salehiyan2025Optimized,
Palety2025Anomaly,Alajlan2025Autonomous,Morshedi2025Combining,
Nguyen2025Enhancing,Shamim2025GAN,Moghaddam2025Generative,
KiranSinghNayak2025Multilayered,Marulli2025Novel}.
Credible generalization requires separation across devices, clients, sites, or
firmware rather than pooled record-level splits; edge-readiness claims further
require measured resource or communication costs on representative settings.

\subsubsection{Cloud, SDN, NFV, and Edge Security}
\label{subsubsec:domain_cloud_sdn}

Cloud, fog, SDN, NFV, and virtualized infrastructures introduce multi-tenancy,
dynamic topology, centralized control, and changing resource configurations.
GAN-based methods address intrusion detection, generated DDoS traffic, anomaly
detection, and cross-domain adaptation
~\cite{DUY2021102367,Senthilkumar2024,QU2024104004,
NAZIM2025101773,10960662,Alabri2026EGDF}.

Evidence in this domain depends particularly on separation across tenants,
providers, controller policies, regions, or network topologies. Deployment
claims additionally require controller load, feature-collection cost,
throughput, latency, or scalability evidence rather than classification
accuracy from a single simulated topology.

\subsubsection{ICS, CPS, and Smart Infrastructure}
\label{subsubsec:domain_ics_cps}

Industrial control, SCADA, smart-grid, advanced-metering, and broader
cyber--physical environments couple cyber observations with physical
processes. GANs support process-anomaly detection, attack-data augmentation,
resilient control, smart-grid intrusion detection, and synthetic industrial
data
~\cite{AGARWAL2024115603,SIDDIQUE2025100281,11200162,
pr13092885,electronics15010235}, including adversarial formulations for
cyber--physical resilience
~\cite{Bhutani2025SADGAN,Saini2026CyResWadc}.

Here, statistical similarity is insufficient if generated observations violate
control logic, process modes, temporal dependence, or physical constraints.
Chronological holdouts, unseen operating regimes, detection delay, process
consequences, and testbed or hardware-in-the-loop validation therefore provide
stronger evidence than random record accuracy alone.

\subsubsection{Wireless, Mobile, and Vehicular Security}
\label{subsubsec:domain_wireless_vehicle}

Wireless, mobile, UAV, 5G/6G, and vehicular systems combine mobility, channel
variation, dynamic topology, intermittent connectivity, and stringent timing
requirements. GAN applications include wireless intrusion detection, UAV
protection, 5G anomaly detection and recovery, vehicular IDS, and
physical-layer authentication
~\cite{YOO2024104073,10902040,11075744,11550383,UMAR2026102009},
with newer work extending to vehicular fog and intelligent-transportation
defense~\cite{Lilhore2026VehicularGAN,Narayan2026ZIDS}.

Evaluation should separate devices, vehicles, routes, channels, or mobility
conditions across development and testing. In safety- or latency-sensitive
settings, detection delay, computational overhead, packet-loss tolerance, and
effects on communication or control functions are more informative than
classification accuracy alone.

\subsubsection{Web, Email, Financial, and Social Security}
\label{subsubsec:domain_web_financial}

GAN applications include phishing augmentation, malicious-email detection,
fraudulent-transaction synthesis, deceptive-review detection, and generative
analysis of financial or social behavior
~\cite{MELISARAHEBI2025103745,9667381,10941350,10342719,
11077152,KHAN202489}, including graph-generative approaches for highly
imbalanced fraud detection
~\cite{Girish2026GraphGANFraud,Zhu2026BADGAN}.

The wider financial-security evidence spans transaction synthesis,
credit-card fraud detection, imbalance mitigation, self-attention, graph, and
hybrid GAN pipelines
~\cite{10246252,Meng2023Detecting,Cheah2023Enhancing,
Aftabi2023Fraud,Almarshad2023Generative,Strelcenia2023Improving,
Alshawi2023Utilizing,Sharma2023Wasserstein,Wang2024Research,
Zhu2025Intelligent,Xie2025GAN,Ge2025Gated,
Selvarajan2025Generative,Pathak2025Improving,Aldakheel2026Hybrid,
Shees2026Zero}. Because accounts, merchants, senders, campaigns, or devices
can recur across observations, chronological and entity-disjoint partitions
are important safeguards against leakage. Precision--recall performance,
controlled false-positive burden, analyst or financial cost, privacy, and
subgroup utility are more informative than aggregate accuracy alone.

\subsubsection{Identity, Biometrics, and Authentication}
\label{subsubsec:domain_identity}

GANs support presentation-attack generation, liveness detection, biometric
augmentation, synthetic-face detection, iris-attack analysis, and
physical-layer authentication~\cite{s21030699,e23081089,app12031454,app13020816,
math11194160,UMAR2026102009,Tynymbayev2026SBDP}.

The principal validity boundary is subject independence, supplemented where
relevant by cross-sensor, cross-instrument, and unseen-attack evaluation.
Biometric synthesis also requires attention to identity memorization,
linkage, sensitive-attribute leakage, and subgroup performance; realistic
samples alone do not establish robust or equitable authentication defense.

\subsubsection{Cross-Domain Cybersecurity}
\label{subsubsec:domain_cross}

This category is reserved for publications whose principal contribution is
evaluated across substantively different cybersecurity environments or
addresses cross-cutting functions such as generative cyber-threat intelligence
~\cite{10257196,HENNA2026114911}. It is not a residual category for poorly
specified applications. Evidence of cross-domain capability requires
identifiable evaluation in each environment; pooling several datasets and
randomly splitting the combined records does not demonstrate transferability
across domains.

Taken together, D3 demonstrates broad application reach but substantial
concentration: network/enterprise, IoT/IIoT/edge, and web/email/financial/social
security account for 72.0\% of the corpus. Domain prevalence should therefore
be distinguished from domain maturity or generalizability, which are examined
through the D3$\times$D5 evidence-maturity analysis in
Section~\ref{subsec:cross_dimensional_analysis}.

\subsection{Dimension 4: Threat or Data Condition--What Risk or Limitation Is Addressed?}
\label{subsec:dim4_threat}

\begin{figure}[!ht]
\centering
\includegraphics[width=\linewidth]{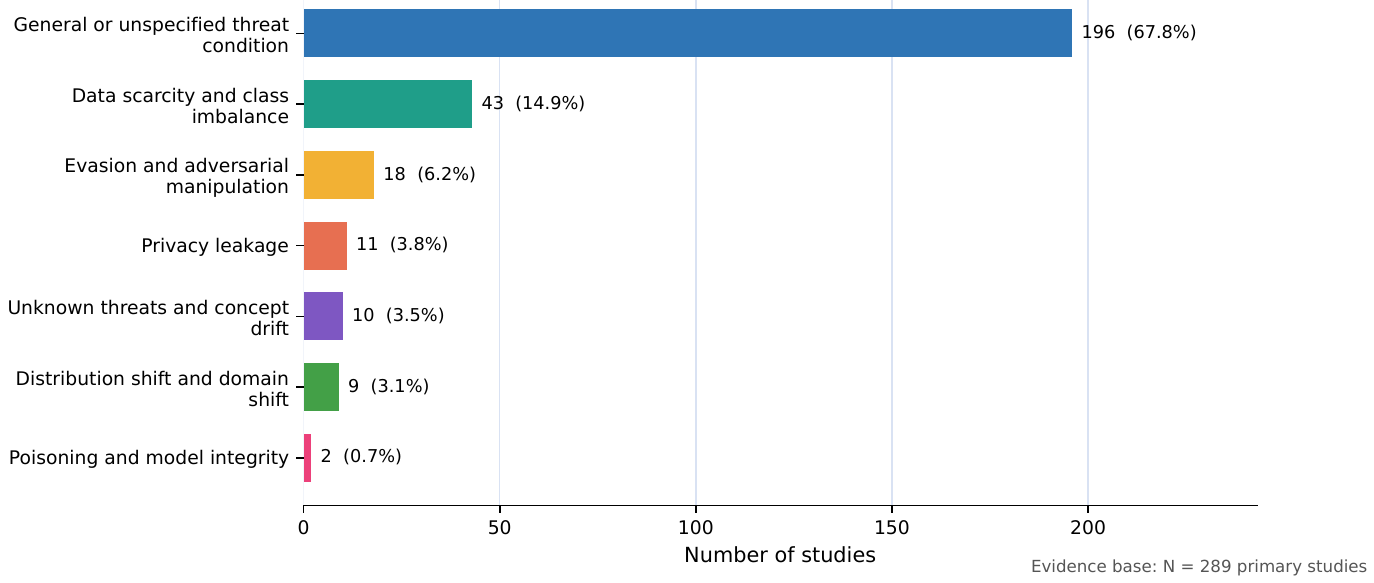}
\caption{\textbf{Primary threat and data-condition distribution (\(N=289\)).}
Each included publication contributes one primary D4 code; attacker-driven
threats are distinguished from non-adversarial data limitations.}
\label{fig:dim4_threat}
\end{figure}

D4 identifies the primary threat or data condition motivating the defense.
Figure~\ref{fig:dim4_threat} reveals a substantial threat-model specificity gap:
196 publications (67.8\%) use a general or unspecified primary condition,
whereas only 93/289 (32.2\%) are anchored to a more explicit threat or data
limitation. Data scarcity and class imbalance account for 43 publications
(14.9\%), followed by evasion and adversarial manipulation (18, 6.2\%),
privacy leakage (11, 3.8\%), unknown threats and concept drift (10, 3.5\%),
distribution and domain shift (9, 3.1\%), and poisoning and model integrity
(2, 0.7\%).

\subsubsection{T1: Data Scarcity and Class Imbalance}
\label{subsubsec:threat_scarcity}

Data scarcity and class imbalance arise when rare attacks, limited incident
sharing, weak labels, or high annotation costs leave defensive models with
insufficient representation of important classes. Although not themselves
attacks, these conditions can produce security failures when aggregate
performance conceals weak minority-class detection.

GAN-based responses include standard, conditional, tabular, graph-based, and
Wasserstein augmentation
~\cite{ANDRESINI2021108,SRIVASTAVA2023103432,ZHAO2024104005,
electronics14204068}, with more recent work addressing class-controlled NIDS,
hybrid conditional generation, financial fraud, Spectre attacks, and
smart-metering security
~\cite{Yang2025CEGAN,Yao2025CWGANGP,math14081264,
Girish2026GraphGANFraud,Priya2026SpectreCGAN,electronics15010235}.
Evidence for T1 is credible only when synthetic training data improve
performance on untouched real observations under leakage-safe partitions and
appropriate imbalance-management baselines; generated sample quantity or
aggregate accuracy alone is insufficient.

\subsubsection{T2: Evasion and Adversarial Manipulation}
\label{subsubsec:threat_evasion}

Evasion occurs when malicious inputs are modified at inference time to mimic
benign behavior or cross a detector's decision boundary. GANs can generate
functionally constrained adversarial traffic, malware, signals, or system
states for defensive stress testing and hardening.

Applications include robust autoencoding, adversarial retraining, ensemble
protection, cyber deception, malware hardening, and post-attack recovery
~\cite{9522215,SARIKAYA2023103483,10437529,QIN2024110655,
11075744,AlEssa2026PEAdversarialGAN}. Attack-oriented systems that generate
evasive IDS traffic or malware contribute threat-model evidence unless their
outputs are incorporated into an implemented and evaluated countermeasure
~\cite{ZHAO2021128,10017127,electronics14030574,Ahmad2025}.
Robustness claims therefore depend on explicit attacker knowledge, manipulable
features, attack budget, operational constraints, transferability, and
evaluation beyond attacks seen during defensive training.

\subsubsection{T3: Poisoning and Model Integrity}
\label{subsubsec:threat_poisoning}

Poisoning targets training data, labels, gradients, client updates, or model
parameters to degrade availability or induce targeted behavior such as
backdoors. It is rare as the primary D4 condition (2/289 publications), although
poisoning-related mechanisms and attacker attributes occur more broadly in the
literature.

GANs may generate poisoned examples, model malicious-update distributions, or
support detection, retraining, and recovery. Related work addresses label
poisoning, federated NIDS, model poisoning, backdoors, and GAN-generated
poisoning
~\cite{HALLAJI2023110384,ZHANG2022154,10574838,
electronics14030406,10945280,11408066}. Credible evaluation must identify the
compromised component, attack objective, poisoning level or malicious-client
fraction, clean utility, attack success, and mitigation persistence. Accidental
label noise is not treated as T3 unless an adversarial integrity compromise is
explicitly modeled.

\subsubsection{T4: Unknown Threats and Concept Drift}
\label{subsubsec:threat_zeroday}

T4 captures unknown, previously unseen, zero-day, open-set, or evolving threat
conditions, including concept drift when the defensive problem is explicitly
framed around changing attack behavior. GAN-based approaches include normality
modeling, reconstruction-based detection, synthetic enrichment, generative
representation learning, and zero-shot generation
~\cite{10604482,BOPPANA2023105805,11416973,11424016,11478946}.

A zero-day claim requires structural exclusion from training, such as a
held-out attack family, later time period, external dataset, unseen malware
lineage, new device population, or formal open-set protocol. Random
closed-set splitting does not establish unknown-threat capability, and
thresholds or hyperparameters should not be tuned using labels from the
excluded target condition.

\subsubsection{T5: Distribution Shift and Domain Shift}
\label{subsubsec:threat_shift}

Distribution and domain shift occur when source and target environments differ
across organizations, networks, devices, software versions, collection periods,
or operational settings. GAN-based responses include adversarial alignment,
target-like generation, few-shot transfer, continual augmentation, and
federated adaptation
~\cite{LI2022102856,MA2023719,BHARDWAJ2024103769,
DUY2024101397,NAZIM2025101773,10.1145/3714334.3714385}.

Credible T5 evidence requires an explicit source--target or chronological
boundary, transparent use of target labels, and comparison against source-only
and relevant non-GAN adaptation baselines. A random split of one dataset does
not demonstrate domain adaptation.

\subsubsection{T6: Privacy Leakage}
\label{subsubsec:threat_privacy}

Cybersecurity data may expose sensitive information about users, devices,
organizations, vulnerabilities, medical conditions, or network structure,
while collaborative learning can leak membership, attributes, gradients, or
local-data properties. GAN-based defenses include synthetic-data sharing,
federated generation, differential privacy, privacy-aware clustering, and
inversion-resistant collaboration
~\cite{TABASSUM2022299,Feizi2024,10437148,Zhao2025,Ullah2026,Hamza2026},
with distributed extensions in fog, IoT, IoMT, and industrial settings
~\cite{Lei2025FogGAN,Nguyen2025FederatedWGAN,Durga2026TinyGAN,
11569888}.

Neither synthetic data nor federation constitutes privacy evidence by itself.
A defensible claim must specify the protected information, attacker access,
implemented leakage or inference test, and any formal privacy mechanism, while
interpreting privacy jointly with defensive utility and communication cost.

\subsubsection{T7: General or Unspecified Threat Conditions}
\label{subsubsec:threat_general}

The largest D4 category comprises publications that evaluate a cybersecurity
defense without identifying a more specific primary attacker, shift, privacy
risk, or data limitation. These publications remain informative about the
reported defensive task, but their evidence does not by itself establish
zero-day capability, adversarial robustness, adaptivity, privacy protection,
or resilience to distributional change. Benchmark-centered intrusion and
anomaly evaluation remains prominent in prior syntheses
~\cite{10187144,ALAJLAN20242053,Alauthman2026,SAEED2026128978}.
When the complete full text did not support a more specific condition, the
publication remained in T7 rather than being assigned an inferred threat model.

Taken together, D4 shows that explicit threat-model specification remains a
minority practice in the corpus: 196/289 publications (67.8\%) are coded as
general or unspecified. Attacker knowledge, capability, attack stage,
objective, budget, and adaptivity are retained as secondary attributes rather
than additional primary D4 categories. The relationship between threat/data
condition and defensive function is examined directly in
Section~\ref{subsec:cross_dimensional_analysis}.

\subsection{Dimension 5: Evidence and Deployment Context--Under What Conditions Is the Defense Validated?}
\label{subsec:dim5_evidence}

\begin{figure}[!ht]
\centering
\includegraphics[width=\linewidth]{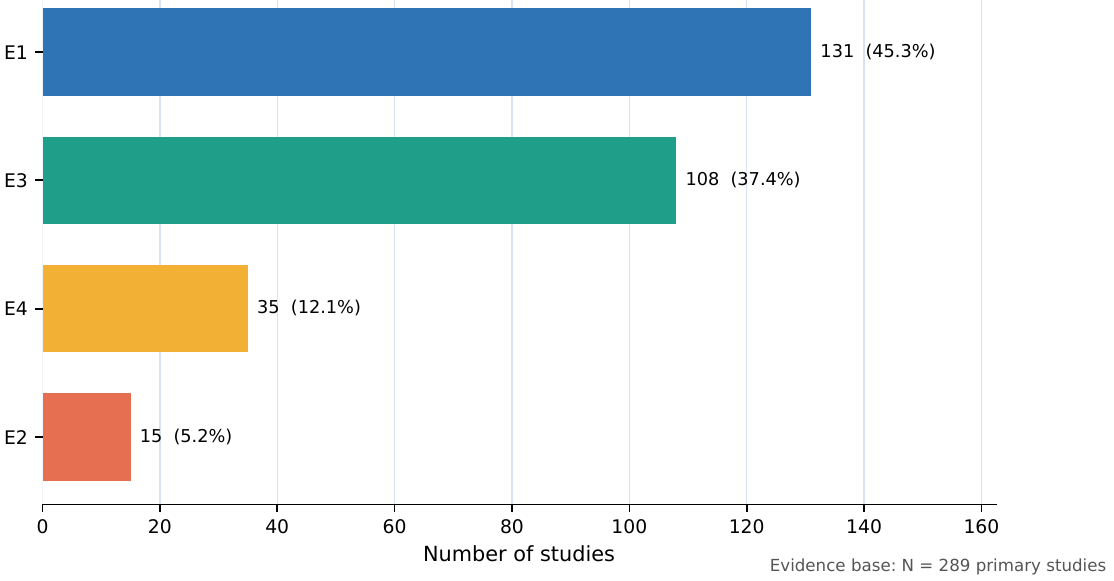}
\caption{\textbf{Primary evidence-maturity distribution (\(N=289\)).}
D5 records the strongest demonstrated validation setting from E1
(offline proof of concept) to E5 (sustained real-world operational deployment).}
\label{fig:dim5_evidence}
\end{figure}

D5 records the strongest demonstrated validation setting supporting each
publication's principal defensive claim. Figure~\ref{fig:dim5_evidence} shows
that 131 publications are classified as E1 (45.3\%), 15 as E2 (5.2\%),
108 as E3 (37.4\%), and 35 as E4 (12.1\%); no publication reaches E5.
Thus, 146/289 publications (50.5\%) remain within offline E1--E2 validation,
while the remainder extend to external, adversarial, temporal, or
deployment-oriented settings without demonstrating sustained operational use.

D5 describes \emph{validation maturity}, not overall methodological quality.
A well-controlled E1 or E2 study may provide more credible evidence than a
poorly executed E4 prototype. Explicit minimum criteria and evidence locations
were therefore used for each assignment, with the full coding checklist
preserved in Supplementary Workbook~S1
~\cite{Ndayipfukamiye2026_GANProtocol}.

\subsubsection{E1: Offline Single-Dataset Proof of Concept}
\label{subsubsec:evidence_e1}

E1 denotes offline evaluation on a single dataset, typically using a
within-dataset partition
~\cite{ANDRESINI2021108,app13042576,SRIVASTAVA2023103432,
ZHAO2024104005}. Such evidence establishes technical feasibility only within
the evaluated data boundary. Leakage-safe partitioning remains essential where
flows, sessions, hosts, malware families, devices, subjects, or adjacent
temporal observations are dependent
~\cite{Dube2024,LAYEGHY2024103689}. E1 does not by itself establish external
generalization, zero-day capability, adaptive robustness, or deployment
readiness.

\subsubsection{E2: Strengthened Offline Validation}
\label{subsubsec:evidence_e2}

E2 remains offline but introduces controls that materially strengthen internal
validity. Qualification requires at least two of the following: evaluation on
three or more public datasets, repeated independent runs with uncertainty
estimates, matched ablations isolating the GAN contribution, or strong
non-GAN baselines addressing the same defensive function
~\cite{electronics14204068,ZOUHRI2025107882,Khalil2026,s26030757}.
Additional classifiers or metrics alone do not establish E2. For synthetic-data
studies, fidelity, diversity, duplication, privacy risk, and
train-on-synthetic--test-on-real utility remain relevant
~\cite{11393807}. E2 strengthens confidence in an offline effect but does not
establish external transfer or deployment feasibility.

\subsubsection{E3: External, Temporal, Open-Set, or Adversarial Validation}
\label{subsubsec:evidence_e3}

E3 introduces a validation boundary not represented in ordinary training.
Qualifying evidence includes meaningful cross-dataset or cross-domain transfer,
chronological evaluation, family- or device-disjoint testing, structurally
open-set evaluation, or adversarial testing with attacks excluded from
defensive training
~\cite{BHARDWAJ2024103769,ROSHAN2024103853,
YUMLEMBAM2024102529,NAZIM2025101773,11007173,10988602}.
Target labels should not influence model selection or thresholding unless the
setting is explicitly transductive or semi-supervised. E3 therefore supports
external, temporal, open-set, or adversarial validity, but not operational
readiness by itself.

\subsubsection{E4: Deployment-Oriented Testbed, Distributed, or Resource-Aware Validation}
\label{subsubsec:evidence_e4}

E4 requires direct deployment-oriented evidence. Qualifying settings include
representative edge or embedded hardware with measured resource cost;
federated or distributed learning with communication and client-heterogeneity
evidence; cyber--physical testbeds or digital twins with process-level
outcomes; or end-to-end pipeline measurements
~\cite{s26030757,TABASSUM2022299,9863068,10908853,10437148,
11007173,10879553,11569888,10988602,11038927,11037504}.
Labels such as \emph{federated}, \emph{edge}, \emph{lightweight}, or
\emph{real-time} are insufficient without corresponding measurements.
Representative examples include fog-enabled federation, local Wasserstein
augmentation, lightweight wireless-sensor detection, federated TinyGANs, and
GAN-supported SDN defense
~\cite{Lei2025FogGAN,Nguyen2025FederatedWGAN,
Devi2024LightweightDCGAN,Durga2026TinyGAN,AlSarray2025GANCLSTO}.
E4 demonstrates deployment-oriented feasibility rather than sustained
operational use.

\subsubsection{E5: Sustained Real-World Operational Deployment}
\label{subsubsec:evidence_e5}

E5 is reserved for defenses evaluated within sustained real-world or pilot
operational workflows using continuously arriving data, analyst or response
integration, online updating, alert-workload measures, or recovery and rollback
evidence. Offline replay, proprietary data, edge execution, or low laboratory
latency alone do not qualify. Relevant outcomes include alert burden,
detection and response time, service disruption, drift, fallback behavior, and
intervention success; explainability claims additionally require evidence of
fidelity, stability, or user usefulness
~\cite{ZOUHRI2025107882,11250986,MELISARAHEBI2025103745}.

No publication in the \(N=289\) corpus met E5. This should be interpreted as
an absence of demonstrated sustained operational evidence in the peer-reviewed
corpus, not as evidence that GAN-based defenses have never been operationally
deployed. Industry, proprietary, or classified deployments may fall outside the
review's eligibility and retrieval boundaries.

Across all D5 levels, evidence maturity is complemented by separate credibility
attributes covering split integrity, synthetic-data fidelity and duplication,
privacy, computational and communication cost, and reproducibility
~\cite{Dube2024,LAYEGHY2024103689,11393807,s26030757,
doi:10.1126/science.359.6377.725,10778911}. These attributes are assessed
independently of D5 so that validation setting is not conflated with study
quality.

\subsection{Defensive Function--Domain Evidence Map}
\label{subsec:study_evidence_map}

Table~\ref{tab:d1_d3_evidence_map} cross-maps the \(N=289\) included
publications by primary defensive function (D1) and cybersecurity domain (D3).
Because each publication contributes one primary code to each dimension, the
matrix provides a non-overlapping descriptive view of where detection,
augmentation, and mitigation-oriented functions are concentrated. Secondary
functions and domains remain available in Supplementary Workbook~S1
~\cite{Ndayipfukamiye2026_GANProtocol}.

\begin{table*}[htbp]
\centering
\caption{Publication-level evidence map across defensive function (D1) and
cybersecurity domain (D3), \(N=289\).}
\label{tab:d1_d3_evidence_map}
\scriptsize

\begin{tabular}{
>{\raggedright\arraybackslash}l
*{9}{>{\centering\arraybackslash}c}
>{\centering\arraybackslash}c}
\toprule
\textbf{Defensive function} &
\rotatebox{60}{\textbf{C1 Network}} &
\rotatebox{60}{\textbf{C2 Malware}} &
\rotatebox{60}{\textbf{C3 IoT/IIoT}} &
\rotatebox{60}{\textbf{C4 Cloud/SDN}} &
\rotatebox{60}{\textbf{C5 ICS/CPS}} &
\rotatebox{60}{\textbf{C6 Wireless/mobile}} &
\rotatebox{60}{\textbf{C7 Web/financial}} &
\rotatebox{60}{\textbf{C8 Identity}} &
\rotatebox{60}{\textbf{C9 Cross-domain}} &
\textbf{Total} \\
\midrule

\textbf{F1} Data augmentation and class balancing
& 24 & 2 & 3 & 1 & 0 & 1 & 10 & 1 & 2 & \textbf{44} \\

\textbf{F2} Adversarial hardening and robustness
& 7 & 8 & 0 & 0 & 0 & 0 & 0 & 0 & 6 & \textbf{21} \\

\textbf{F3} Anomaly and threat detection
& 95 & 12 & 32 & 7 & 15 & 7 & 20 & 4 & 8 & \textbf{200} \\

\textbf{F4} Privacy-aware synthesis and collaboration
& 8 & 0 & 4 & 0 & 0 & 0 & 1 & 0 & 1 & \textbf{14} \\

\textbf{F5} Domain adaptation
& 2 & 1 & 0 & 0 & 0 & 0 & 0 & 0 & 0 & \textbf{3} \\

\textbf{F6} Deception, testing, or recovery
& 1 & 0 & 1 & 0 & 2 & 1 & 0 & 0 & 2 & \textbf{7} \\

\midrule
\textbf{Total}
& \textbf{137} & \textbf{23} & \textbf{40} & \textbf{8}
& \textbf{17} & \textbf{9} & \textbf{31} & \textbf{5}
& \textbf{19} & \textbf{289} \\
\bottomrule
\end{tabular}
\footnotesize
\textit{Note}: Values are mutually exclusive primary-code counts. D1 and D3 labels follow Table~\ref{tab:taxonomy_overview}; complete publication membership, secondary tags, and coding evidence are preserved in Supplementary Workbook~S1~\cite{Ndayipfukamiye2026_GANProtocol}.
\end{table*}

Detection is the largest single defensive function in each of the nine domains,
with network-oriented detection alone accounting for 95/289 publications
(32.9\%). Mitigation-oriented functions (F2, F4, F5, and F6) are proportionally
most visible in cross-domain cybersecurity (9/19, 47.4\%) and malware/endpoint
security (9/23, 39.1\%), whereas augmentation is concentrated primarily in
network and web/financial settings. These patterns describe the distribution
of research activity rather than comparative effectiveness and complement the
prespecified cross-dimensional association analyses in
Section~\ref{subsec:cross_dimensional_analysis}.


\subsection{Cross-Dimensional Structure of the Design Space}
\label{subsec:cross_dimensional_analysis}

Marginal distributions describe the prevalence of individual taxonomy
categories, but they do not show whether defensive functions, GAN
architectures, cybersecurity domains, threat conditions, and evidence-maturity
levels co-occur more often than expected from their respective margins. For
dimensions \(X\) and \(Y\), the observed count in cell \((a,b)\) was defined as
\begin{equation}
n_{ab}^{XY}
=
\sum_{i=1}^{N}
\mathbb{I}\!\left(X_i=a \land Y_i=b\right),
\qquad N=289 .
\label{eq:cross_dimensional_count}
\end{equation}
The corresponding expected count under independence was
\(E_{ab}=n_{a\cdot}n_{\cdot b}/N\). Association magnitude was summarized using
Cram\'er's
\(V=\sqrt{\chi^2/[N\min(r-1,c-1)]}\), where \(r\) and \(c\) denote the
numbers of non-empty row and column categories. Because sparse contingency
tables can upward-bias ordinary Cram\'er's \(V\), we additionally report the
finite-sample bias-corrected coefficient \(V_{\mathrm{bc}}\).

Because several expected cell counts were small, asymptotic \(p\)-values were
not used as the principal association diagnostic. Instead, each observed
Pearson \(\chi^2\) statistic was compared with a fixed-margin permutation
distribution obtained by permuting one dimension across publications while
preserving both marginal distributions. We used \(B=50{,}000\) permutations
with random seed 42 and calculated the Monte Carlo probability using the
plus-one correction:
\[
p_{\mathrm{perm}}
=
\frac{
1+\sum_{b=1}^{B}
\mathbb{I}\!\left(
\chi^2_b \geq \chi^2_{\mathrm{obs}}
\right)}
{B+1}.
\]
Adjusted Pearson residuals were examined to identify the cells contributing
most strongly to each global association. These residuals are interpreted
descriptively rather than as separate multiplicity-adjusted hypothesis tests.

The analyses characterize structural association among the publications
identified by this review. They do not imply population representativeness,
causal dependence, or comparative effectiveness of one architecture, domain,
or defensive function over another.

\begin{figure}[!ht]
\centering
\includegraphics[width=\linewidth]
{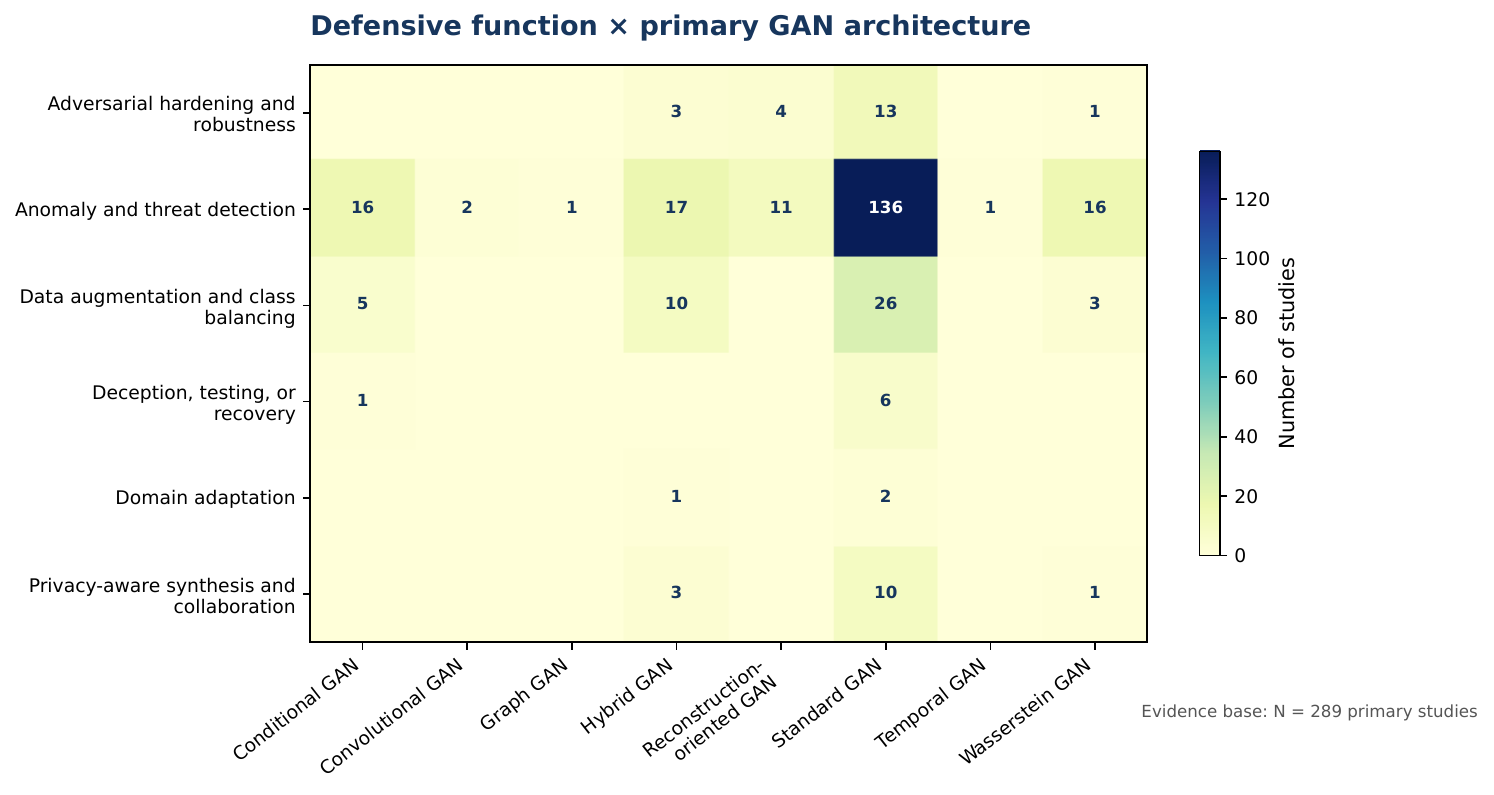}
\caption{Defensive function (D1) by primary GAN architecture (D2),
\(N=289\). Counts represent one primary category per included publication. Descriptive and bias-corrected Cram\'er's \(V\), together with fixed-margin permutation testing, are used because the contingency table is sparse.}
\label{fig:function_architecture_heatmap}
\end{figure}

Figure~\ref{fig:function_architecture_heatmap} shows little evidence of a
systematic D1$\times$D2 association
(\(\chi^2=28.81\), \(V=0.141\), \(V_{\mathrm{bc}}=0.000\),
\(p_{\mathrm{perm}}=0.540\)). The largest cell is anomaly and threat detection
with standard GANs (\(n=136\)), whereas approximately 133.6 publications would
be expected in this cell from the dominant row and column margins alone. The
permutation result therefore provides no evidence that GAN architectures are
concentrated systematically within particular defensive functions beyond those
margins. This finding does not establish that architectures are technically
equivalent or equally effective; rather, it shows that architecture labels
alone do not sharply differentiate the defensive roles assigned in the
reviewed corpus.

\begin{figure}[!ht]
\centering
\includegraphics[width=\linewidth]
{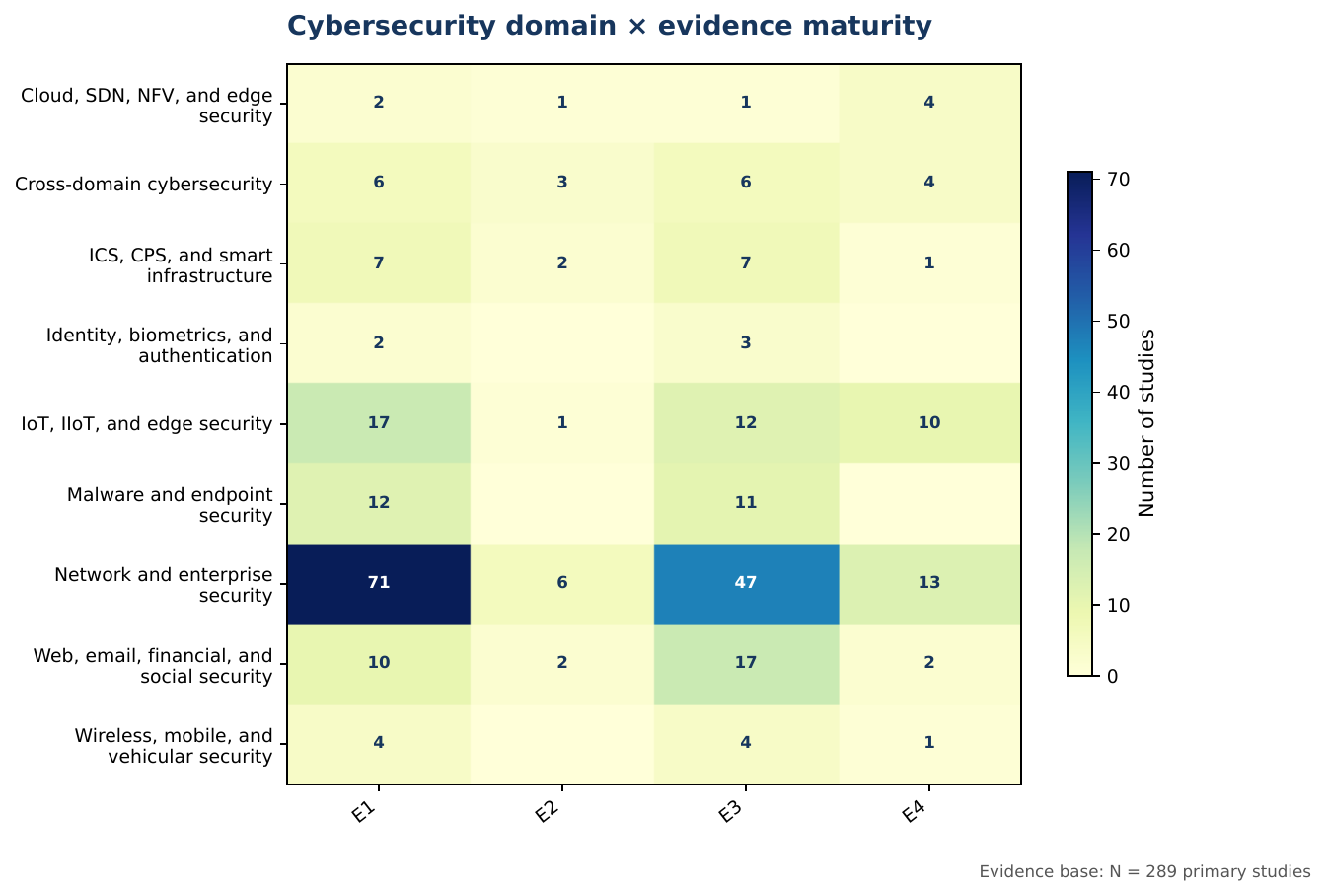}
\caption{Cybersecurity domain (D3) by evidence maturity (D5), \(N=289\). E5 is omitted from the displayed contingency matrix and association calculation because no included publication reached sustained operational validation. Its zero occupancy is retained as a substantive finding rather than treated as missing data.}
\label{fig:domain_evidence_heatmap}
\end{figure}

Figure~\ref{fig:domain_evidence_heatmap} indicates a modest
\(D3\times D5\) association
(\(\chi^2=41.73\), \(V=0.219\), \(V_{\mathrm{bc}}=0.143\),
\(p_{\mathrm{perm}}=0.0186\); 95\% Monte Carlo interval:
\([0.0174,\,0.0198]\)). Under the final \(B=50{,}000\) fixed-margin
permutation analysis, the Monte Carlo estimate and its interval lie below
the nominal \(\alpha=0.05\) threshold; however, because three prespecified
cross-dimensional associations were examined, the result does not meet the
Bonferroni sensitivity threshold of \(\alpha=0.0167\) and is therefore
interpreted as descriptive rather than multiplicity-robust evidence.

The largest positive adjusted residuals occur for Cloud/SDN/NFV/edge security
at E4 (\(n_{\mathrm{obs}}=4\), \(n_{\mathrm{exp}}=0.97\),
\(r_{\mathrm{adj}}=3.33\)) and IoT/IIoT/edge security at E4
(\(n_{\mathrm{obs}}=10\), \(n_{\mathrm{exp}}=4.84\),
\(r_{\mathrm{adj}}=2.69\)). Network and enterprise security is comparatively
concentrated at E1 (\(n_{\mathrm{obs}}=71\),
\(n_{\mathrm{exp}}=62.10\), \(r_{\mathrm{adj}}=2.11\)).

These departures identify domains in which the assembled evidence has
progressed relatively further toward external, testbed, resource-aware, or
distributed-system evaluation. They do not demonstrate that the domains
themselves are intrinsically more mature or that their GAN-based defenses are
more effective. E5 was excluded from the association calculation because its
marginal count is zero; the absence of any publication demonstrating sustained
operational validation remains a substantive review finding.

\begin{figure}[!ht]
\centering
\includegraphics[width=\linewidth]
{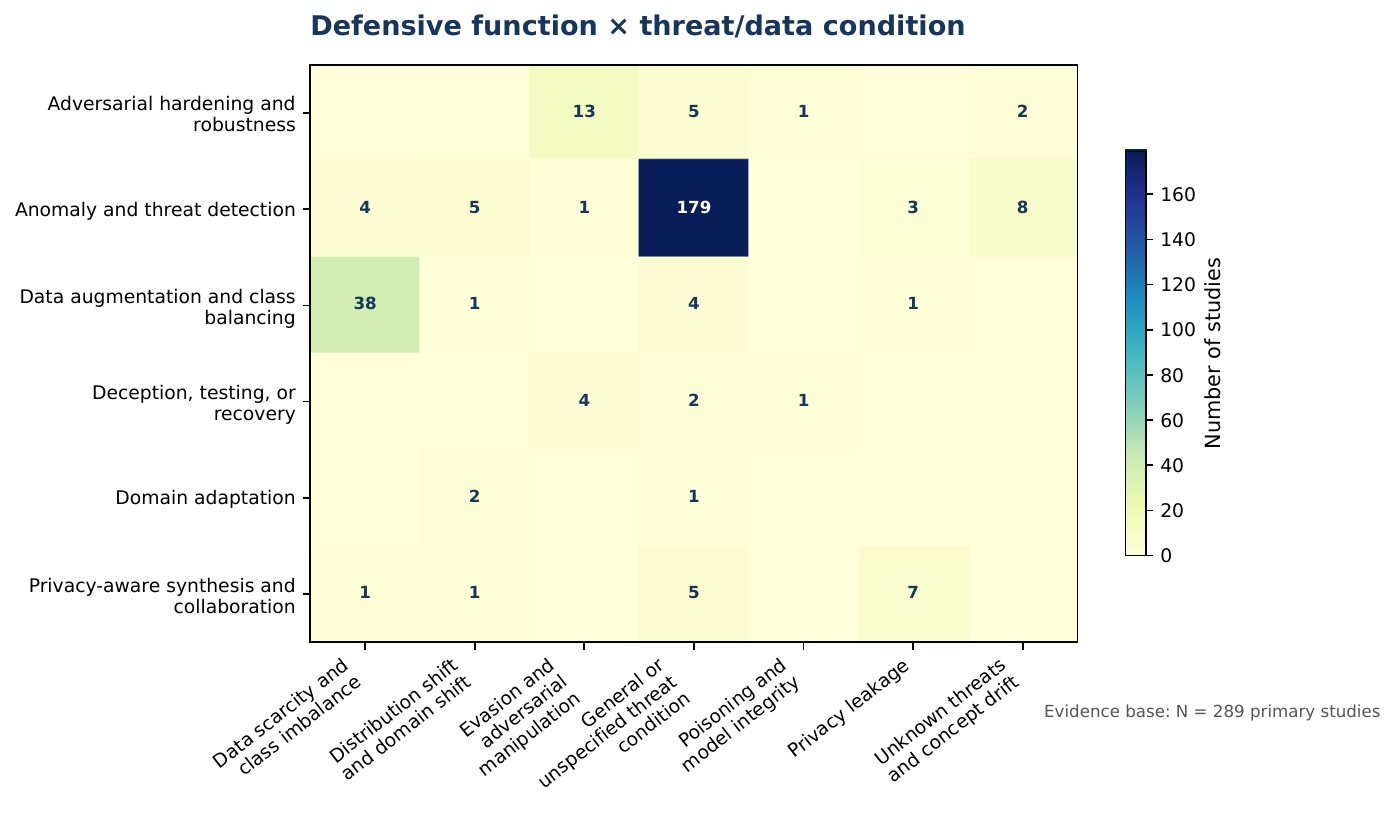}
\caption{Defensive function (D1) by primary threat or data condition (D4), \(N=289\). The concentration reflects both coherent function--condition alignments and the predominance of general or unspecified threat conditions among detection publications.}
\label{fig:function_threat_heatmap}
\end{figure}

The strongest cross-dimensional structure occurs between defensive function
and primary threat or data condition
(Figure~\ref{fig:function_threat_heatmap}:
\(\chi^2=526.92\), \(V=0.604\), \(V_{\mathrm{bc}}=0.592\),
\(p_{\mathrm{perm}}<2.0\times10^{-5}\)); none of the 50,000
fixed-margin permutations produced a statistic as large as the observed
value. The largest positive adjusted residuals correspond to data augmentation
with scarcity or class imbalance
(\(38/44\), \(r_{\mathrm{adj}}=14.47\)),
anomaly and threat detection with a general or unspecified threat condition
(\(179/200\), \(r_{\mathrm{adj}}=11.83\)),
adversarial hardening with evasion or adversarial manipulation
(\(13/21\), \(r_{\mathrm{adj}}=10.96\)), and
privacy-aware synthesis with privacy leakage
(\(7/14\), \(r_{\mathrm{adj}}=9.26\)).
Domain adaptation is also aligned with distribution or domain shift
(\(2/3\)), although this function contains only three publications.

The association therefore reflects two different properties of the corpus. First, several specialized functions are coherently aligned with the security failure they are intended to address. Second, most detection publications are coded under a general or unspecified primary threat condition. The latter pattern limits the extent to which high detection performance can be interpreted as evidence against a clearly defined attacker, distributional change, or operational threat.

To determine whether the global association was produced solely by the
dominant general or unspecified category, we removed that category and
recomputed the \(D1\times D4\) analysis. This removed 196 publications and retained \(n=93\) publications across the six explicit threat or data-condition categories. The association remained strong
(\(\chi^2=185.43\), \(V=0.631\), \(V_{\mathrm{bc}}=0.604\),
\(p_{\mathrm{perm}}<2.0\times10^{-5}\)). The category
``unknown threats and concept drift'' was retained because it is a substantive threat condition rather than a residual or unclear category. Thus, the alignment between specialized defensive functions and explicit problem conditions is not an artifact of the general or unspecified category, although the result remains a structural description of this review corpus rather than a population-level estimate.

The magnitude of the D1$\times$D4 association should nevertheless not be
interpreted as evidence that two fully independent constructs have become empirically coupled. Several category pairs are construct-coupled by design: augmentation and class balancing directly address scarcity or imbalance, adversarial hardening directly addresses evasion or manipulation, and privacy-aware synthesis directly addresses privacy
leakage. Their large residuals therefore provide primarily a coherence
check on the taxonomy rather than an independent substantive discovery.
The less definitionally constrained result is that 179/200 detection
publications fall under a general or unspecified primary threat condition, revealing that most detection studies are not organized around an explicit attacker, distributional shift, privacy failure, or other specific threat condition.

Taken together, the cross-dimensional analyses provide two different
insights. First, GAN architecture is only weakly related to defensive
function after sparse-table correction
($V=0.141$, $V_{\mathrm{bc}}=0.000$), indicating that architecture
labels alone do not organize the defensive design space. Second, D1 and
D4 show strong alignment, but part of that alignment is structurally
expected from the coding definitions. The more consequential finding is
the limited threat specificity of the detection literature: 179/200
detection publications use a general or unspecified primary threat
condition. Architecture selection should therefore begin with the
security failure, attacker or distributional condition, admissible
outcome, and required evidence level rather than with the popularity of a named GAN family.


\subsection{Quality Appraisal and Corpus Sensitivity}
\label{subsec:taxonomy_quality}

Quality appraisal was restricted to the \(n=231\) original experimental
studies. Under the eight-item rubric, 109 studies (47.2\%) were classified
as high quality, 102 (44.2\%) as moderate, and 20 (8.7\%) as low. The
remaining 58 included publications were not appraised because they did not
report an original experiment meeting the experiment-level criteria.

Reporting completeness remained uneven. As detailed in
Section~\ref{sec:empirical_synthesis}, 207/231 studies (89.6\%) reported a
comparator, but only 31 (13.4\%) included an explicit ablation, 16 (6.9\%)
reported repeated runs, 2 (0.9\%) explicitly reported random seeds, and
17 (7.4\%) provided a public author artifact. These indicators describe
reporting support rather than defensive effectiveness; taxonomic prevalence
should therefore not be interpreted as methodological strength.

\subsubsection{Sensitivity to Post-Inclusion Analytical Stratification}
\label{subsec:sensitivity_limited_evidence}

We assessed whether restricting the taxonomy from the full included corpus
(\(N=289\)) to the original experimental subset (\(n=231\)) materially
changed the primary D1--D5 distributions. For each dimension,
Table~\ref{tab:sensitivity} reports the category with the largest absolute
percentage-point difference.

\begin{table}[!ht]
\centering
\caption{Sensitivity of the primary D1--D5 distributions to restriction
to the original experimental subset. For each dimension, the category with
the largest absolute difference is reported.}
\label{tab:sensitivity}
\scriptsize
\setlength{\tabcolsep}{1pt}
\renewcommand{\arraystretch}{0.5}
\begin{tabular}{lp{2.0cm}ccc}
\toprule
\textbf{Dimension} &
\textbf{Category with largest difference} &
\textbf{Full corpus} &
\textbf{Experimental subset} &
\(\boldsymbol{|\Delta|}\) \textbf{(pp)} \\
&
&
\(\boldsymbol{N=289}\) &
\(\boldsymbol{n=231}\) &
\\
\midrule
D1 & Data augmentation and class balancing
   & 44/289 (15.22\%) & 38/231 (16.45\%) & 1.23 \\
\addlinespace
D2 & Standard GAN
   & 193/289 (66.78\%) & 149/231 (64.50\%) & 2.28 \\
\addlinespace
D3 & Web, email, financial, and social security
   & 31/289 (10.73\%) & 23/231 (9.96\%) & 0.77 \\
\addlinespace
D4 & Data scarcity and class imbalance
   & 43/289 (14.88\%) & 39/231 (16.88\%) & 2.00 \\
\addlinespace
D5 & E3: External, temporal, or adversarial validation
   & 108/289 (37.37\%) & 83/231 (35.93\%) & 1.44 \\
\bottomrule
\end{tabular}
\end{table}

The maximum difference was 2.28 percentage points, for standard GANs,
indicating that the principal marginal D1--D5 patterns are descriptively
stable under restriction to the experimental subset. This analysis does not
address possible selection bias from the 410 reports that could not be retrieved.


\subsection{Integrated Taxonomy Contribution and Answers to RQ1--RQ2}
\label{subsec:taxonomy_synthesis}

The taxonomy synthesis reveals three structural patterns. First, defensive
function and threat/data condition are strongly aligned
(\(V=0.604\), \(V_{\mathrm{bc}}=0.592\)), but 179 of 200 detection
publications use a general or unspecified primary threat condition. Second,
architecture and defensive function show little systematic association
(\(V=0.141\), \(V_{\mathrm{bc}}=0.000\),
\(p_{\mathrm{perm}}=0.540\)), indicating that architecture labels alone do
not determine defensive role. Third, evidence maturity remains concentrated
in offline settings (E1: 45.3\%) and reaches E4 in only 35 publications, with
no publication reaching sustained operational validation (E5).

These findings reframe GAN-based cybersecurity defense as a problem of design
alignment rather than architecture selection. The primary implication is that
future work should prioritize: (i) explicit threat modeling with
function-appropriate baselines and ablations; (ii) leakage-resistant,
entity- or time-aware partitions; (iii) reproducible artifact release; and
(iv) deployment-oriented evaluation beyond offline proof-of-concept.

The answer to RQ1 is therefore not a list of GAN architectures: the corpus
comprises six defensively distinct functions implemented through reusable
architectural mechanisms across nine operational domains. The answer to RQ2
is that defensive function is structured much more strongly by the motivating
threat or data condition than by GAN architecture, while evidence maturity
remains domain dependent and stops short of sustained deployment. Together,
D1--D5 provide a problem-first design logic: define the defensive failure and
threat boundary, select a domain-valid generative mechanism, and specify the
evidence level required before claiming effectiveness.


\section{Experimental Evidence and Effectiveness Synthesis}
\label{sec:empirical_synthesis}

This section answers RQ3 using the \(n=231\) included publications that reported
an identifiable original experimental evaluation. The remaining 58 publications
remain part of the \(N=289\) review corpus and contribute to systematic mapping
and qualitative synthesis, but not to experiment-level frequency denominators.

\subsection{Experimental Evidence Base and Reporting Profile}
\label{subsec:experimental_profile}

The experiment-level extraction covered datasets, preprocessing, GAN
configuration, comparators, metrics, validation, implementation, ablation,
repetition, uncertainty, reproducibility artifacts, operational evidence, and
numerical results. Table~\ref{tab:experimental_reporting_profile} summarizes the
principal reporting indicators. Presence denotes that the item was explicitly
identifiable; it does not by itself establish methodological adequacy.

\begin{table}[!t]
\centering
\caption{Experiment-level reporting profile for the \(n=231\) original
experimental studies.}
\label{tab:experimental_reporting_profile}
\scriptsize
\setlength{\tabcolsep}{0.5pt}
\renewcommand{\arraystretch}{0.8}
\begin{tabular}{lrrlrr}
\toprule
\textbf{Indicator} & \textbf{\(n\)} & \textbf{\%} &
\textbf{Indicator} & \textbf{\(n\)} & \textbf{\%} \\
\midrule
Dataset/data source identified & 231 & 100.0 &
Comparator reported & 207 & 89.6 \\
GAN configuration & 230 & 99.6 &
Explicit ablation & 31 & 13.4 \\
Downstream model/baseline information & 230 & 99.6 &
Repeated runs & 16 & 6.9 \\
Evaluation metrics & 220 & 95.2 &
Random seed explicitly reported & 2 & 0.9 \\
Preprocessing & 202 & 87.4 &
Dispersion reported & 52 & 22.5 \\
Numerical result & 194 & 84.0 &
Confidence interval & 9 & 3.9 \\
Hyperparameters & 192 & 83.1 &
Statistical test & 24 & 10.4 \\
Sample size/class balance & 164 & 71.0 &
Public author artifact & 17 & 7.4 \\
Validation protocol & 160 & 69.3 &
Deployment/operational evidence & 61 & 26.4 \\
Implementation detail adequate & 154 & 66.7 &
Hardware reported & 130 & 56.3 \\
Framework/software reported & 127 & 55.0 &
 & & \\
\bottomrule
\end{tabular}
\end{table}

The principal imbalance is therefore between high reporting of model
configuration, metrics, and comparators and much weaker evidence for attribution,
stochastic reliability, reproducibility, and operational validation. This limits
how confidently favorable benchmark results can be attributed specifically to the
GAN component.

\subsection{Datasets, Partitions, and Evaluation Environments}
\label{subsec:datasets_protocols}

Named public or research datasets dominated the experimental evidence
(204/231; 88.3\%); 14 studies (6.1\%) used study-generated or simulated data,
seven (3.0\%) study-collected data, three (1.3\%) mixed sources, two (0.9\%)
private/proprietary data, and one (0.4\%) an unnamed external source.
Dataset use was non-exclusive because individual studies could evaluate multiple
benchmarks. The most frequent were NSL-KDD (50 studies; 21.6\%),
CICIDS2017 (46; 19.9\%), UNSW-NB15 (41; 17.7\%), KDD Cup~99
(26; 11.3\%), BoT-IoT (16; 6.9\%), the European Credit Card Fraud dataset
(15; 6.5\%), and ToN-IoT (13; 5.6\%).

Preprocessing was described in 202/231 studies (87.4\%), sample or class
composition in 164 (71.0\%), and validation protocols in 160 (69.3\%).
These details are consequential because record-level random partitions may place
related flows, devices, malware families, or adjacent time periods across training
and testing. Augmentation claims therefore require generation from training data
only and evaluation on untouched real observations, while zero-day, drift, and
transfer claims require appropriate family-, entity-, time-, device-, or
domain-separated holdouts~\cite{BHARDWAJ2024103769,Riaz2025,Dube2024,
LAYEGHY2024103689}.

\subsection{Metrics and Claim--Outcome Alignment}
\label{subsec:metrics_alignment}

Evaluation remained strongly classification-oriented. Accuracy was reported in
202/231 studies (87.4\%), precision in 173 (74.9\%), recall in 168 (72.7\%),
F1 in 164 (71.0\%), ROC in 67 (29.0\%), AUC in 65 (28.1\%), false-positive
rate in 64 (27.7\%), and detection rate in 56 (24.2\%). Runtime appeared in
22 studies (9.5\%) and FID in 12 (5.2\%). Metric frequencies are non-exclusive.

Metric presence is not equivalent to claim validity. Accuracy alone can obscure
minority-class failure; detection metrics do not establish synthetic-data fidelity
or privacy; and low model complexity does not establish real-time operation.
Outcomes must therefore remain aligned with the claimed function: augmentation
with class-sensitive utility on real test data, hardening with clean and attacked
utility, detection with false-alarm and open-set behavior, privacy with direct
leakage or formal guarantees, and deployment with system-level resource
measures~\cite{10858943,11393807,s26030757}.

\subsection{Baselines, Ablations, and Attribution}
\label{subsec:baseline_ablation}

A comparator was reported in 207/231 experiments (89.6\%), but only 31
(13.4\%) contained an explicit ablation. Thus, comparative reporting was common,
whereas isolation of the GAN's incremental contribution was not. In hybrid
pipelines, changes to classifiers, features, preprocessing, class ratios, optimizers,
or training budgets can account for observed improvements. Credible attribution
therefore requires matched no-GAN and function-appropriate comparisons, with
other major components held constant or explicitly ablated
~\cite{ZHAO2024104005,AlEssa2026PEAdversarialGAN,
ZOUHRI2025107882,HENNA2026114911}.

\subsection{Uncertainty, Repetition, and Statistical Reliability}
\label{subsec:uncertainty}

Repeated runs were explicitly reported in only 16/231 studies (6.9\%) and random
seeds in 2 (0.9\%). Dispersion measures appeared in 52 studies (22.5\%),
confidence intervals in nine (3.9\%), and statistical tests in 24 (10.4\%).
Because GAN optimization, sampling, initialization, and partitioning are stochastic,
the predominance of point estimates limits assessment of result stability and
precludes construction of a common uncertainty-aware effect distribution. This
supports the use of structured SWiM synthesis rather than conventional
meta-analysis.

\subsection{Reproducibility and Operational Evidence}
\label{subsec:experimental_reproducibility}

Implementation transparency was uneven: hyperparameters were reported in
192/231 studies (83.1\%), adequate implementation detail in 154 (66.7\%),
hardware in 130 (56.3\%), and software/framework information in 127 (55.0\%).
Public author artifacts were much rarer: 17 studies (7.4\%) provided at least
one, including code in 16 (6.9\%), author data in four (1.7\%), model/checkpoint
files in two (0.9\%), and configuration or scripts in two (0.9\%).

Deployment- or operationally relevant evidence appeared in 61/231 studies
(26.4\%), but this includes testbed and resource-aware evaluation rather than
sustained field deployment. None of the \(N=289\) included publications reached
E5 sustained operational validation. The literature therefore demonstrates
prototype and bounded external evidence rather than persistent effectiveness
under evolving threats and operational change~\cite{s25010130,s26030757,
10430114,11037504}.

\subsection{Matched Contrasts and Bounded Effectiveness}
\label{subsec:matched_contrasts}

The matched-contrast registry contained 150 candidate comparisons sharing
nominal dataset and metric labels. None satisfied both split and preprocessing
equivalence requirements, and six of the 150 were invalid self-contrasts.
Consequently, no candidate was eligible for pooled comparison, directional
voting, or a common effect estimate. The complete audit is retained in Supplementary Workbook~S1 (Worksheet~18)~\cite{Ndayipfukamiye2026_GANProtocol}.

This restriction avoids treating GAN versus an unbalanced classifier as equivalent to GAN versus conventional resampling, interpreting comparisons between GAN families as a general ``GAN effect,'' or combining substantively different outcomes. Individual studies nevertheless provide bounded evidence that augmentation, attack-specific hardening, and domain adaptation can be useful under their reported conditions~\cite{ZHAO2024104005,Yang2025CEGAN,
AlEssa2026PEAdversarialGAN,11075744,BHARDWAJ2024103769}; these findings do not establish a common effect or universal architectural superiority.

\subsection{Structured Narrative Synthesis by Defensive Function}

Because cross-study pooling was not admissible, Table~7 summarizes the three dominant functions within the \(n=231\) experimental subset. Augmentation (\(n=38\)), detection (\(n=161\)), and hardening (\(n=16\)) jointly account for 215/231 experiments (93.1\%).

\begin{table*}[!ht]
\centering
\caption{Bounded narrative synthesis of the three dominant defensive functions
within the original experimental subset.}
\scriptsize
\setlength{\tabcolsep}{0.5pt}
\renewcommand{\arraystretch}{0.5}
\begin{tabular}{p{2.0cm}p{4.5cm}p{5.5cm}p{5.5cm}}
\toprule
\textbf{Function} &
\textbf{Defensible synthesis} &
\textbf{Principal evidence boundary} &
\textbf{Minimum evaluation requirement} \\
\midrule

Augmentation
(F1; \(n=38\))
&
GAN-generated minority samples can improve class-sensitive performance within
controlled pipelines, but no common effect magnitude is estimable.
&
Changes in classifiers, preprocessing, class ratios, or partitions frequently
prevent attribution specifically to GAN augmentation.
&
Generate from training data only; retain real test data and the same detector;
compare with no augmentation and conventional resampling.
\\
\addlinespace
Detection
(F3; \(n=161\))
&
GAN-based reconstruction or anomaly scoring can support detection under the
evaluated benchmark conditions.
&
The primary threat condition was general or unspecified in 142/161 studies
(88.2\%), limiting inference about unseen, temporal, or open-set threats.
&
Specify the threat model; use appropriate entity-, family-, time-, or
domain-separated evaluation; report class-sensitive errors and false-alarm burden.
\\
\addlinespace
Hardening
(F2; \(n=16\))
&
GAN-generated adversarial examples can improve resistance to evaluated attacks
within stated attacker conditions.
&
Evidence against unseen, transferred, or adaptive attacks remains limited, and
robustness may trade off against clean-data utility.
&
Match attacker knowledge and budget; report clean and attacked utility; test
unseen/adaptive attacks and include a matched no-GAN ablation.
\\
\bottomrule
\end{tabular}
\end{table*}

Across these functions, credible inference depends more on alignment among the
security claim, threat model, partition, comparator, and outcome than on the GAN
family alone. RQ3 therefore supports conditional, task-specific utility under
documented experimental conditions, but neither a common effect size nor
architecture-independent superiority.

\section{Technical Challenges, Reproducibility, and Deployment Barriers}
\label{sec:challenges}

This section addresses \textbf{RQ4}: \textit{What technical, methodological,
reproducibility, trustworthiness, and operational limitations restrict current
GAN-based cybersecurity defenses?} The synthesis integrates author-reported
limitations with the experiment-level audit and D5 evidence-maturity findings.
Five interconnected barrier classes emerge (Table~\ref{tab:verified_barrier_synthesis}):
generative-model validity, data and benchmark validity, evaluation and threat
validity, reproducibility and attribution, and deployment and operational
trustworthiness. An unreported property is recorded as not reported and is not
assumed to be satisfied. Supporting evidence locations are preserved in
Supplementary Workbook~S1~\cite{Ndayipfukamiye2026_GANProtocol}.

\begin{table*}[htbp]
\centering
\caption{Principal barriers to trustworthy and deployable GAN-based
cybersecurity defense.}
\label{tab:verified_barrier_synthesis}
\scriptsize
\setlength{\tabcolsep}{1.2pt}
\renewcommand{\arraystretch}{0.5}
\begin{tabular}{p{2.0cm}p{6.5cm}p{3.5cm}p{5.5cm}}
\toprule
\textbf{Barrier} &
\textbf{Observed evidence pattern} &
\textbf{Consequence} &
\textbf{Required evidence} \\
\midrule

Generative-model validity
&
Training instability, incomplete mode coverage, memorization, and statistically
plausible but semantically or functionally invalid samples
~\cite{NIPS2014_f033ed80,pmlr-v70-arjovsky17a,NIPS2017_892c3b1c,
Feizi2024,11393807}.
&
Synthetic volume or favorable runs may not improve valid threat coverage.
&
Repeated runs, sensitivity analysis, diversity and memorization checks,
domain constraints, and untouched real-data evaluation.
\\
\addlinespace

Data and benchmark validity
&
Public benchmarks may contain duplication, outdated attacks, simulation
artifacts, unrealistic class priors, or leakage-prone partitions
~\cite{Dube2024,LAYEGHY2024103689}.
&
Within-dataset performance may overstate transferable defensive capability.
&
Versioned provenance, duplicate analysis, pre-generation partitioning, and
entity-, family-, device-, or time-disjoint testing.
\\
\addlinespace

Evaluation and threat validity
&
Although 207/231 experiments reported a comparator, only 31/231 included an
explicit ablation; robustness, zero-day, privacy, and adaptation claims also
depend on attacker and target-data assumptions
~\cite{ROSHAN2024103853,YUMLEMBAM2024102529,
BHARDWAJ2024103769,AlEssa2026PEAdversarialGAN}.
&
The claim may exceed the experiment or the observed gain may not be attributable
to the GAN.
&
Matched no-GAN ablations, function-specific alternatives, explicit threat
models, structurally unseen conditions, and claim-appropriate outcomes.
\\
\addlinespace

Reproducibility and attribution
&
Only 16/231 experiments reported repeated runs, 2/231 explicitly reported
random seeds, and 17/231 provided a public author artifact
~\cite{doi:10.1126/science.359.6377.725,10778911}.
&
Results may be difficult to reconstruct or distinguish from favorable stochastic
variation.
&
Exact data and split identifiers, complete configurations and environments,
repeated runs, uncertainty, executable artifacts, and component ablation.
\\
\addlinespace

Deployment and trustworthiness
&
Only 61/231 experiments reported deployment-related evidence, while no
publication in the \(N=289\) corpus reached E5 sustained operational validation;
privacy, drift, safety, human workload, and rollback were also unevenly assessed
~\cite{s26030757,10430114,s25010130,11250986,11037504}.
&
Offline predictive success may not translate to continuous, distributed,
safety-sensitive, or human-supervised operation.
&
End-to-end resource measurements, adaptive and longitudinal testing, privacy
evaluation, human-centered outcomes, rollback, and sustained operational
evidence.
\\
\bottomrule
\end{tabular}
\end{table*}

\subsection{Model, Data, and Evaluation Validity}
\label{subsec:technical_data_limitations}

\subsubsection{Generative Stability and Security-Valid Data}
\label{subsubsec:generative_validity_barrier}

GAN-based defenses inherit stochastic adversarial-optimization problems,
including oscillation, non-convergence, initialization sensitivity, and
generator--discriminator imbalance
~\cite{NIPS2014_f033ed80,pmlr-v70-arjovsky17a,NIPS2017_892c3b1c}.
Stability-oriented objectives can improve optimization, but stable losses do
not establish mode coverage, absence of memorization, or defensive utility;
privacy mechanisms may further introduce utility--privacy--convergence
trade-offs~\cite{Feizi2024}.

Security-valid synthesis is also domain dependent. Network data should preserve
protocol and session logic, malware transformations the claimed functionality,
temporal and graph data their structural dependencies, and industrial or
cyber--physical observations their physical feasibility. Stronger evaluations
therefore combine multidimensional fidelity, domain constraints, diversity and
duplication checks, and utility on untouched real data
~\cite{11393807,electronics15010235,11200162,pr13092885}.

\subsubsection{Benchmark Provenance and Leakage}
\label{subsubsec:benchmark_validity_barrier}

A generator cannot correct defects already embedded in its source data.
Duplicated observations, outdated attacks, simulation artifacts, incomplete
provenance, and unrealistic class distributions may instead be reproduced or
amplified. Analyses of CIC-IDS2017 and comparisons between public NIDS
benchmarks and production traffic demonstrate that dataset construction,
preprocessing, and representation materially affect conclusions
~\cite{Dube2024,LAYEGHY2024103689}.

Leakage arises when generation precedes partitioning, preprocessing uses test
information, or dependent flows, hosts, malware families, devices, accounts,
subjects, or time periods cross the split. Partitioning must therefore precede
generator fitting and reflect the dependency structure of the domain. Public
reproducibility and operational representativeness remain distinct properties:
a benchmark may be fully accessible yet operationally unrealistic.

\subsubsection{Claim, Threat, and Attribution Validity}
\label{subsubsec:evaluation_validity_barrier}

The central validity problem is mismatch between the security claim and the
implemented experiment. Closed-set classification does not establish zero-day
detection; attacks used during training do not demonstrate adaptive robustness;
federation does not establish privacy; and classifier-only latency does not
demonstrate real-time operation.

Robustness depends on attacker knowledge, perturbation constraints,
transferability, functionality, and the defended classifier
~\cite{ROSHAN2024103853,YUMLEMBAM2024102529,
AlEssa2026PEAdversarialGAN}. Adaptation similarly depends on a meaningful
source--target boundary and transparent use of target labels
~\cite{BHARDWAJ2024103769}, while distributed systems introduce poisoning and
model-integrity risks rather than eliminating the attacker surface
~\cite{10574838,11408066,11007173}.

Attribution is therefore as important as final performance. The high comparator
rate in the experimental subset (207/231, 89.6\%) contrasts with the much lower
explicit-ablation rate (31/231, 13.4\%). For augmentation, hybrid systems,
privacy, adaptation, and hardening, credible incremental GAN value requires
matched no-GAN or function-specific alternatives and outcomes aligned with the
stated claim.

\subsection{Reproducibility and Evidence Attribution}
\label{subsec:reproducibility_gaps}

\subsubsection{Artifact and Configuration Completeness}
\label{subsubsec:artifact_completeness}

Reconstructing a GAN-based defense requires more than a repository link. The
dataset version and split, preprocessing, generator and discriminator
configuration, losses, optimizer and update schedule, augmentation policy,
thresholds, random seeds, hardware, software environment, and evaluation
pipeline jointly determine the reported result. Cybersecurity-AI reproduction
studies document failures caused by version conflict, software and hardware
incompatibility, and tool obsolescence~\cite{10778911}; broader evidence
likewise emphasizes inspectable methods and artifacts
~\cite{doi:10.1126/science.359.6377.725}.

Generated samples, checkpoints, split identifiers, and evaluation scripts are
particularly useful for auditing fidelity, leakage, thresholding, and metric
definitions. Where raw data cannot be released, provenance, feature
definitions, class distributions, split rules, and a reproducible sanitized or
public validation setting remain important substitutes.

\subsubsection{Stochasticity and Hybrid Attribution}
\label{subsubsec:stochasticity_attribution}

GAN performance varies with initialization, data order, update ratios,
augmentation levels, and stopping conditions. Yet only 16/231 experiments
reported repeated runs and only 2/231 explicitly reported random seeds,
limiting assessment of whether an observed gain is stable rather than a
favorable training trajectory.

Hybrid systems create an additional attribution burden because adversarial
generation is frequently combined with feature selection, autoencoders,
attention, transformers, graph networks, federation, optimization, or
explainability
~\cite{ZOUHRI2025107882,HAMOUDA2024101149,Khalil2026,
10.1145/3700838.3700871,11555542}. Component-level ablation is therefore
essential when attributing a pipeline improvement specifically to the GAN.
Energy-aware cross-dataset benchmarking illustrates that stronger reporting
and artifact release are feasible~\cite{s26030757}.

Computational reproduction and external replication should also remain
distinct: reproducing the same implementation verifies reconstruction, whereas
testing another environment, population, attack family, dataset, or time
period evaluates generality.

\subsection{Deployment and Operational Trustworthiness}
\label{subsec:deployment_barriers}

\subsubsection{Resources and Distributed-System Feasibility}
\label{subsubsec:resource_scalability_barriers}

GAN-based defenses incur cost across generator training, synthetic-data
production, downstream retraining, preprocessing, inference, communication,
and updating. Measuring only the final classifier can therefore understate
deployment burden. Energy-aware benchmarking shows that architecture rankings
may change when predictive utility is considered jointly with energy
~\cite{s26030757}, while accelerated Industry~4.0 pipelines provide useful
hardware evidence without establishing feasibility across constrained devices
~\cite{s25010130}.

Lightweight, fog, and federated approaches respond to these constraints
~\cite{Devi2024LightweightDCGAN,Durga2026TinyGAN,Lei2025FogGAN,
Nguyen2025FederatedWGAN,AlSarray2025GANCLSTO}, but labels such as
``lightweight,'' ``edge,'' or ``federated'' require measured support.
Representative hardware, end-to-end latency, memory, energy, transmitted
volume, client heterogeneity, non-IID severity, participation, convergence, and
retraining cost are particularly relevant; distributed systems must also
account for malicious or unreliable clients
~\cite{10908853,10879553,10988602,11037504}.

\subsubsection{Adaptive Threats, Privacy, and Human Oversight}
\label{subsubsec:adaptive_privacy_human_barriers}

Operational environments evolve as attackers adapt, software changes, devices
enter and leave, and class priors drift. One-shot benchmark evaluation therefore
does not establish longitudinal robustness. More credible evidence incorporates
future-window or cross-domain evaluation, unseen and defense-aware attacks,
drift detection, secure retraining, poisoning resistance, update validation,
and rollback
~\cite{ROSHAN2024103853,CHEN2024103987,10574838,11408066,
11007173}.

Synthetic generation and federation can reduce direct raw-data exchange but are
not intrinsically private. Privacy claims require an explicit attacker,
implemented membership, inversion, reconstruction, attribute, or
gradient-leakage test, or a formal guarantee with transparent accounting
~\cite{Feizi2024,TABASSUM2022299,10437148,11569888}. Utility and
communication cost must be interpreted alongside privacy.

Human-facing deployment requires equally direct evidence. Explainable or
human-in-the-loop GAN defenses are increasingly reported
~\cite{ZOUHRI2025107882,10902040,11250986,HENNA2026114911,
MELISARAHEBI2025103745}, but explanation availability alone does not establish
operational usefulness. Faithfulness, stability, actionability, alert burden,
investigation time, and authority over automated response require evaluation
with representative users.

\subsubsection{Safety, Operational Validation, and Dual Use}
\label{subsubsec:operational_validation_barriers}

Statistical realism is insufficient where generated observations or defensive
actions can violate protocols, executable behavior, physical processes, or
safety constraints. Autonomous recovery, industrial testbeds, constrained
generation, and accelerated monitoring provide stronger deployment-oriented
evidence
~\cite{10430114,pr13092885,electronics15010235,s25010130,
Bhutani2025SADGAN}, but hardware execution or replayed traffic does not by
itself constitute sustained operational validation.

Operational evidence should instead document the deployment environment,
duration, workload, alert handling, human interaction, model updating,
failures, maintenance, rollback, service interruption, and recovery. The
absence of E5 evidence in the \(N=289\) corpus therefore marks an important
boundary between deployment-oriented feasibility and demonstrated sustained
operation.

GAN-based security is also dual use: mechanisms developed for hardening,
testing, or deception can support evasion, malware mutation, phishing, or
automated probing. Responsible release consequently requires
capability-specific risk assessment, controlled evaluation environments,
appropriate access restrictions, misuse monitoring, and disclosure procedures
rather than generic ethical statements
~\cite{COPPOLINO2025129406,Shafik2025,Ibrar2025}.

Taken together, these barriers are cumulative rather than independent. Stable generation on unrepresentative data does not establish generalization;
leakage-prone performance does not establish defensive effectiveness;
reproducible offline results do not establish deployment readiness; and
resource-efficient systems remain operationally incomplete when privacy,
adaptive robustness, safety, or human control are untested. Section~\ref{sec:discussion_agenda} translates these evidence gaps into research and engineering priorities.

\section{Discussion and Future Research Agenda}
\label{sec:discussion_agenda}

\subsection{Evidence-Centered Interpretation}

The central finding of this review is that the defensive value of GANs depends
less on architecture labels than on alignment among defensive function,
threat/data condition, domain, evaluation protocol, and evidence maturity.
The cross-dimensional results reinforce this interpretation: architecture
and defensive function show little systematic association
(\(V=0.141\), \(V_{\mathrm{bc}}=0.000\)), whereas defensive function and
threat/data condition show strong alignment
(\(V=0.604\), \(V_{\mathrm{bc}}=0.592\)); the latter magnitude is
interpreted cautiously because several D1--D4 category pairs are
construct-coupled by definition. Domain and evidence maturity are also
related, although more modestly
(\(V_{\mathrm{bc}}=0.143\)).

The most consequential imbalance is therefore not architectural but
evidential. Threat detection accounts for 200/289 publications (69.2\%), yet
179 of those 200 use a general or unspecified primary threat condition.
Likewise, 131/289 publications (45.3\%) remain at E1 offline proof-of-concept
maturity and none reaches E5 sustained operational validation. The literature
therefore demonstrates substantial algorithmic activity, but considerably less
evidence of threat-specific, longitudinal, and operational validity.

The experiment-level synthesis reinforces this interpretation. Although
207/231 experimental studies (89.6\%) reported a comparator, explicit ablation
was reported in only 31 (13.4\%), repeated runs in 16 (6.9\%), explicit random
seeds in 2 (0.9\%), and public author artifacts in 17 (7.4\%). Moreover, none
of the 150 candidate cross-study contrasts satisfied the predefined
split-and-preprocessing equivalence requirements. Consequently, the evidence
supports conditional, task-specific GAN utility rather than a transferable
claim of architectural superiority.

\subsection{Implications for Research and Practice}

For researchers, the target claim should be specified before model selection:
incremental GAN value, full-system performance, robustness to a defined
attacker, privacy benefit, domain transfer, or deployment feasibility require
different comparators and outcomes. Matched no-GAN ablations, leakage-resistant
partitions, strong function-specific alternatives, repeated runs, uncertainty
reporting, and claim-appropriate evaluation should therefore be treated as
core experimental requirements rather than optional additions.

For practitioners and system owners, adoption should depend on the similarity
between published evidence and the intended data, threat, hardware, and
operational workflow. External or temporal validation, false-positive burden,
failure modes, privacy and integrity testing, resource cost, monitoring,
retraining, rollback, and human authorization are more informative for
deployment decisions than isolated benchmark accuracy.

For benchmark designers, reproducible evaluation requires versioned provenance,
duplicate and contamination analysis, entity- or time-aware partitions, fixed
preprocessing manifests, class-sensitive metrics, and independent test
conditions. Benchmarks should distinguish within-environment performance from
structurally held-out, temporal, and cross-domain generalization.

\subsection{Research Priorities}

\subsubsection{Security-Valid Data and Threat-Valid Evaluation}

Future work should evaluate whether generated observations preserve the
properties that determine security validity: protocol and session logic for
traffic, executable behavior for malware, temporal and relational structure
for sequences and graphs, and physical feasibility for industrial or vehicular
systems. Fidelity should therefore be assessed jointly with constraint
violations, rare-mode coverage, duplication, memorization, and utility on
untouched real data
~\cite{11393807,Dube2024,LAYEGHY2024103689}.

Threat evaluation should likewise distinguish seen, unseen, transferable,
adaptive, and functionality-constrained attacks and explicitly report attacker
knowledge, access, objective, and budget. Zero-day, adaptation, and drift claims
require structural separation across attack families, time periods, devices,
networks, organizations, or other meaningful deployment boundaries rather than
random partitions.

\subsubsection{Reproducible and Comparative Evidence}

Reproducibility should extend beyond descriptive reporting to a complete
experiment manifest covering dataset version, preprocessing and partitions,
leakage controls, architecture and hyperparameters, seeds, independent runs,
hardware/software environments, and artifact locations. The low frequencies of
public artifacts, repeated runs, and explicit seeds observed in the
experimental subset indicate that stochastic reliability and reproducibility
remain substantive evidence limitations.

Comparative evaluation should also test whether a GAN is necessary. Class
weighting, conventional oversampling, autoencoders, one-class models, diffusion
models, flows, autoregressive models, secure aggregation, and differentially
private learning should be evaluated under matched data, tuning, hardware,
threat, and outcome conditions. A more complex GAN pipeline is justified only
when it provides reproducible incremental value over credible alternatives.

\subsubsection{Joint Trustworthiness and Deployment Evaluation}

Utility, robustness, privacy, calibration, and resource efficiency should be
evaluated jointly rather than as independent claims. Privacy-oriented studies
require direct leakage testing or formal guarantees; federated systems require
evidence on non-IID severity, participation, malicious-client assumptions,
convergence, communication, and local computation; and deployment studies
require end-to-end latency, memory, energy, throughput, maintenance, monitoring,
and rollback
~\cite{10184476,Feizi2024,10908853,s26030757}.

Progress beyond E4 additionally requires longitudinal evidence from realistic
operational workflows. Future evaluations should document deployment duration,
changing workloads, drift, analyst interaction, update behavior, failures,
fallback procedures, and recovery rather than treating hardware execution or
laboratory latency as equivalent to operational validation.

\subsubsection{Responsible Development and Dual Use}

GAN-based security research remains inherently dual use. Mechanisms developed
for augmentation, hardening, testing, or deception may also enable evasion,
malware mutation, phishing, or automated probing. Reproducibility should
therefore be combined with proportionate capability-risk controls: benign
evaluation manifests, metric definitions, and sanitized scripts can generally
be released openly, whereas high-capability malicious artifacts, sensitive
operational data, or deployable evasion systems may require controlled access,
isolated evaluation, staged release, monitoring, or independent governance
~\cite{Shafik2025,Ibrar2025}. Responsible-release practices should be reported
as part of the experimental and deployment context rather than treated as
generic ethical statements.

\subsection{Limitations of the Review}
\label{subsec:review_limitations}

Six limitations qualify the findings. 
First, Web of Science, grey literature, preprints, and systematic forward-citation searching were not included, so indexing and publication-selection bias cannot be excluded, particularly given that Scopus was the primary discovery pathway for 203/289 included publications (70.2\%).

Second, 410/718 database-pathway reports (57.1\%) were not retrieved, all from Scopus; because systematic author contact and formal interlibrary-loan retrieval were not undertaken, retrieval-related selection bias remains possible, especially for recent Scopus-indexed publications.

Third, some legacy searches lacked exact execution dates, while Scopus
expansion and SpringerLink validation were revision-stage amendments documented retrospectively in the OSF record~\cite{Ndayipfukamiye2026_GANProtocol}. SpringerLink capture was limited to 1,000 of 3,314 displayed results, and the
exact single-query execution form of the retained ScienceDirect search could
not be reconstructed, introducing additional source-specific coverage and
reproducibility uncertainty.

Fourth, screening and coding were reviewer-led; calibration and verification improved consistency but did not constitute complete independent duplicate assessment. Assigning one primary D1--D5 category per publication also necessarily simplifies hybrid systems despite secondary tagging.

Fifth, experiment-level synthesis was restricted to the \(n=231\) publications reporting an original experimental evaluation. Restricting the taxonomy from the full \(N=289\) corpus to this subset changed no primary D1--D5 prevalence estimate by more than 2.28 percentage points, although this sensitivity analysis does not address the 410 non-retrieved reports. 

Sixth, heterogeneous outcomes, evaluation protocols, and limited uncertainty reporting precluded meta-analysis, while the 2026 stratum covers only January--June and has had less time for replication and post-publication scrutiny.

Accordingly, the findings support conditional, problem-specific uses of GANs rather than universal architectural superiority.

\section{Conclusion}
\label{sec:conclusion}

This review synthesizes \(N=289\) included full-text publications on
GAN-based cybersecurity defense published from January~2021 through
June~2026, of which \(n=231\) reported an identifiable original experimental
evaluation. The five-dimensional taxonomy integrates defensive function,
GAN architecture, cybersecurity domain, threat/data condition, and evidence
maturity, while the experiment-level synthesis distinguishes directly
supported evidence from claims that cannot be compared reliably across
heterogeneous experimental settings.

At the corpus level, research is concentrated in anomaly and threat detection
(200 publications, 69.2\%), standard GAN formulations (193, 66.8\%), and
network/enterprise security (137, 47.4\%). Mitigation-oriented functions,
including adversarial hardening, privacy-aware collaboration, domain
adaptation, and deception, account for only 45 publications (15.6\%).
Architecture and defensive function show little systematic association
(\(\chi^2=28.81\), \(V=0.141\), \(V_{\mathrm{bc}}=0.000\),
\(p_{\mathrm{perm}}=0.540\)), indicating that named GAN families do not
sharply determine defensive role. This is an important structural result:
architecture labels alone provide limited guidance about the security
function for which a GAN is used.

Defensive function and threat/data condition show substantially stronger
alignment (\(\chi^2=526.92\), \(V=0.604\),
\(V_{\mathrm{bc}}=0.592\),
\(p_{\mathrm{perm}}<2.0\times10^{-5}\)). This magnitude should, however, be
interpreted cautiously because several D1--D4 relationships are partly
construct-coupled by definition, including augmentation with data scarcity,
adversarial hardening with evasion or manipulation, and privacy-aware
synthesis with privacy leakage. The association therefore serves partly as a
coherence check on the taxonomy rather than as evidence of a fully independent
empirical relationship between defensive function and threat/data condition.
More substantively, 179/200 detection publications (89.5\%) use a general or
unspecified primary threat condition, revealing limited threat specificity
within the dominant detection literature. Evidence maturity is likewise
limited: 131 publications (45.3\%) remain at E1 offline proof-of-concept
maturity, and none reaches E5 sustained operational validation
(Section~\ref{subsubsec:evidence_e5}).

The \(n=231\) experimental subset reveals a similar imbalance between benchmark
reporting and evidential rigor. Although 207/231 studies (89.6\%) reported a
comparator, only 31 (13.4\%) included an explicit ablation, 16 (6.9\%)
reported repeated runs, 2 (0.9\%) explicitly reported random seeds, and
17 (7.4\%) provided a public author artifact. Moreover, none of the 150
candidate cross-study contrasts satisfied the predefined requirements for an
admissible pooled comparison. The available evidence therefore supports
conditional, task-specific uses of GANs under reported experimental
conditions, but does not support a common cross-study effect magnitude or a
universally superior GAN architecture.

Progress requires threat-explicit evaluation, leakage-resistant and
time- or entity-aware partitions, matched no-GAN and conventional baselines,
component-level ablations, repeated-run uncertainty reporting, reusable
artifacts, and deployment-oriented validation. GANs should therefore be
treated as conditional mechanisms for clearly defined defensive failures
rather than as default architectures whose novelty, popularity, or benchmark
performance substitutes for evidence. The complete review corpus, screening
audit, taxonomy codes, experiment-level extractions, methodological-quality
appraisal, matched-contrast audit, and validation materials are available at
\url{https://doi.org/10.17605/OSF.IO/WQHSP}.

\section*{Acknowledgments}
The authors gratefully acknowledge the guidance and collegial support received during this research, as well as the institutional resources that facilitated its completion.

\section*{Funding}
This work was partly supported by the European Celtic+ and Swedish Vinnova project (C2022/1-3), ``CISSAN--Collective Intelligence Supported by Security Aware Nodes.''

\section*{Data Availability Statement}
The retrospective protocol/revision snapshot, search strategy, screening and eligibility decisions, taxonomy coding registry, methodological-quality appraisal, experiment-level evidence register, matched-contrast audit, validation checks, and supporting figure data underlying this review are openly available via the Open Science Framework at~\url{https://doi.org/10.17605/OSF.IO/WQHSP}. This record serves as the persistent archive for the supplementary materials and is not presented as prospective preregistration.

\section*{Code Availability Statement}
Record normalization, consistency checks, descriptive statistics, cross-tabulation, and figure-source calculations were performed using spreadsheet formulas within Supplementary Workbook~S1 archived at~\url{https://doi.org/10.17605/OSF.IO/WQHSP}.
The permutation-based diagnostics reported in Section~\ref{subsec:cross_dimensional_analysis}
(\(\chi^{2}\), Cram\'er's \(V\), and fixed-margin permutation tests for
Figs.~10--12) were computed in Python using SciPy and NumPy, with random
seed 42 and \(B=50{,}000\) permutations for each test. The archived script,
\texttt{cross\_dimensional\_tests\_v2.py}, reproduces the statistics reported in
Section~\ref{subsec:cross_dimensional_analysis}.

\section*{Competing Interests}
The authors declare no competing interests.

\section*{Author Contributions}
T.N. contributed to conceptualization, methodology, investigation, data curation, formal analysis, visualization, and writing--original draft. J.D. contributed to methodology, validation, supervision, and writing--review and editing. D.S.S. and A.G.P. contributed to investigation, evidence verification, and writing--review and editing. H.N. contributed to supervision, project administration, funding acquisition, and writing--review and editing. All authors reviewed and approved the final manuscript.


\section*{Ethics Approval}
Ethics approval was not required because this study involved no human participants, animal subjects, interventions, or collection of sensitive personal data. All synthesized evidence was derived from previously published, peer-reviewed literature.

\appendix
\begingroup
\renewcommand{\thetable}{T.\arabic{table}} 
\setcounter{table}{0} 
\renewcommand{\thefigure}{F.\arabic{figure}} 
\setcounter{figure}{0} 

\section{Source-Specific Search Strategy}
\label{app:search_strings}

The search syntax was adapted to the field restrictions and query capabilities
of each source. Table~\ref{tab:source_search_log} reports the exact final queries
and the number of records retained in the corresponding exports. Export filenames, filters, source-specific search logs, and available search/export-date evidence are preserved in Supplementary Workbook~S1 at~\url{https://doi.org/10.17605/OSF.IO/WQHSP}. Exact execution dates that could not be recovered from the retained legacy records are marked unavailable rather than inferred.

\begin{table*}[htbp]
\centering
\caption{Electronic sources, retained records, and search logic.}
\label{tab:source_search_log}
\scriptsize
\setlength{\tabcolsep}{0.5pt}
\renewcommand{\arraystretch}{0.8}
\begin{tabular}{p{2.0cm}cp{15.0cm}}
\toprule
\textbf{Source} & \textbf{Captured} & \textbf{Source-adapted query} \\
\midrule
\textbf{IEEE Xplore} & 925 & \texttt{("Generative Adversarial Network" OR "GAN" OR "GANs" OR "WGAN" OR "WGAN-GP" OR "CGAN" OR "DCGAN" OR "GANomaly" OR "AnoGAN" OR "MalGAN") AND ("adversarial defense" OR "adversarial attack" OR "adversarial training" OR "data augmentation" OR "synthetic data") AND ("cybersecurity" OR "intrusion detection" OR "IDS" OR "malware" OR "IoT" OR "phishing" OR "fraud" OR "anomaly detection")} \\
\addlinespace
\textbf{SCOPUS} & 2,954 & \texttt{TITLE-ABS-KEY ( ( "generative adversarial network*" OR "GAN-based" OR "GAN enabled" OR CGAN OR DCGAN OR WGAN OR "WGAN-GP" OR ACGAN OR AnoGAN OR GANomaly OR MalGAN OR IDSGAN OR TimeGAN OR CycleGAN ) AND ( cybersecurity OR "cyber security" OR "cyber defense" OR "cyber defence" OR "network security" OR "information security" OR "intrusion detection" OR "network intrusion" OR malware OR ransomware OR botnet OR phishing OR DDoS OR "denial of service" OR "IoT security" OR "internet of things security" OR IIoT OR "industrial control system*" OR SCADA OR "cyber physical system*" OR "cyber-physical system*" OR "smart grid security" OR "cloud security" OR "edge security" OR "wireless security" OR "vehicular network security" OR "fraud detection" OR "biometric security" OR "authentication security" ) AND ( "data augmentation" OR "class balancing" OR "class imbalance" OR "synthetic data" OR "adversarial training" OR hardening OR robustness OR "intrusion detection" OR "anomaly detection" OR "malware detection" OR "threat detection" OR "attack detection" OR "fraud detection" OR "privacy preserving" OR "privacy-preserving" OR "federated learning" OR "domain adaptation" OR deception OR honeypot* OR recovery OR mitigation OR "spoof detection" OR "anti-spoofing" ) ) AND PUBYEAR > 2020 AND PUBYEAR < 2027 AND ( LIMIT-TO ( LANGUAGE , "English" ) ) AND ( LIMIT-TO ( DOCTYPE , "cp" ) OR LIMIT-TO ( DOCTYPE , "ar" ) OR LIMIT-TO ( DOCTYPE , "ch" ) ) AND PUBYEAR > 2020 AND PUBYEAR < 2027} \\
\addlinespace
\textbf{ACM Digital Library} & 1,063 & \texttt{("Generative Adversarial Network" OR "GAN" OR "GANs" OR "WGAN" OR "WGAN-GP" OR "CGAN" OR "DCGAN") AND ("cybersecurity" OR "intrusion detection" OR "IDS" OR "malware" OR "IoT" OR "phishing" OR "fraud" OR "anomaly detection")} \\
\addlinespace
\textbf{ScienceDirect} & 5,177 & \texttt{ALL("Generative Adversarial Network" OR "GAN" OR "GANs" OR "WGAN" OR "WGAN-GP" OR "CGAN" OR "DCGAN" OR "GANomaly") AND ALL("adversarial defense" OR "adversarial attack" OR "adversarial training" OR "data augmentation" OR "synthetic data") AND ALL("cybersecurity" OR "intrusion detection" OR "IDS" OR "malware" OR "IoT" OR "phishing" OR "fraud" OR "anomaly detection")} \\
\addlinespace
\textbf{MDPI} & 1,278 & \texttt{("Generative Adversarial Network" OR "GAN" OR "GANs" OR "WGAN" OR "CGAN" OR "DCGAN") AND ("cybersecurity" OR "intrusion detection" OR "IDS" OR "malware" OR "IoT" OR "phishing" OR "fraud" OR "anomaly detection")} \\
\addlinespace
\textbf{SpringerLink} & 1,000 & \texttt{("Generative Adversarial Network" OR "GAN" OR "GANs" OR "WGAN" OR "WGAN-GP" OR "CGAN" OR "DCGAN") AND ("cybersecurity" OR "intrusion detection" OR "IDS" OR "malware" OR "IoT" OR "phishing" OR "fraud" OR "anomaly detection")} \\
\midrule
\textbf{Total captured} & \textbf{12,397} & Records initially captured; 167 SpringerLink query artifacts were removed before the validated \textbf{12,230-record} deduplication input. \\
\bottomrule
\end{tabular}
\end{table*}


\section{Supplementary Evidence and Reproducibility Materials}
\label{app:supplementary_index}

Supplementary Workbook~S1
(DOI: \url{https://doi.org/10.17605/OSF.IO/WQHSP})
preserves the record-level audit trail and analytical data underlying the
review. Its contents are summarized in Table~\ref{tab:supplementary_index}.

\begin{table*}[htbp]
\centering
\caption{Contents of Supplementary Workbook~S1.}
\label{tab:supplementary_index}
\scriptsize
\setlength{\tabcolsep}{0.5pt}
\renewcommand{\arraystretch}{0.6}
\begin{tabular}{lp{14.0cm}}
\toprule
\textbf{Worksheet} & \textbf{Content} \\
\midrule

\texttt{00\_README}
&
Workbook structure, denominator rules, analytical strata, worksheet roles,
and OSF information.
\\

\texttt{01\_DASHBOARD}
&
Graphical summary of identification, publication, taxonomy, evidence-maturity,
quality, and experiment-level results.
\\

\texttt{02\_SEARCH\_STRATEGY}
&
Database-specific searches, captured records, validation removals, temporal
exclusions, and source provenance.
\\

\texttt{03--08\_RAW\_*}
&
Raw bibliographic exports from IEEE Xplore, ACM Digital Library,
ScienceDirect, MDPI, SpringerLink, and Scopus.
\\

\texttt{09\_PRISMA}
&
PRISMA 2020 count register for database/register and other-method pathways,
ending in \(N=289\) included full-text publications and the post-inclusion
split into \(n=231\) original experimental studies and 58 other included
publications.
\\

\texttt{09A\_SCREENING\_CALIBRATION}
&
Calibration records and consistency checks supporting the title--abstract
and eligibility screening process.
\\

\texttt{10\_DEDUPLICATION}
&
Record-level duplicate/overlap decisions, canonical records, matching rules,
and verification notes.
\\

\texttt{11\_TA\_SCREENING}
&
Title--abstract screening decisions, codes, pathways, and supporting evidence.
\\

\texttt{12\_FT\_ELIGIBILITY}
&
Retrieval and eligibility ledger for 741 reports
(718 database-pathway reports plus 23 unique backward-reference reports),
distinguishing reports not retrieved, full-text exclusions, and the
\(N=289\) included publications.
\\

\texttt{13\_FT\_EXCLUSIONS}
&
The 42 full-text exclusions with terminal reason, decision code, bibliographic
identity, and supporting rationale.
\\

\texttt{14\_INCLUDED\_STUDIES}
&
Authoritative \(N=289\) included-publication registry with bibliographic
metadata, discovery provenance, and post-inclusion experimental status.
\\

\texttt{15\_QUALITY\_Q1\_Q8}
&
Q1--Q8 appraisal for the \(n=231\) original experimental studies, including
criterion scores and quality categories; the remaining 58 publications are
marked not applicable.
\\

\texttt{16\_TAXONOMY\_D1\_D5}
&
Primary and secondary D1--D5 taxonomy assignments with coding evidence and
study-level audit fields.
\\

\texttt{17\_EVIDENCE\_EXPERIMENTS}
&
Experiment-level register for \(n=231\), covering datasets, preprocessing,
GAN configuration, comparators, metrics, validation, ablation, repetitions,
uncertainty, implementation, artifacts, deployment, and numerical results.
\\

\texttt{18\_MATCHED\_CONTRASTS}
&
Audit of 150 candidate contrasts for dataset, split, preprocessing, metric,
and domain equivalence; none met the predefined admissibility conditions.
\\

\texttt{19\_CODEBOOK}
&
Operational definitions for corpus strata, D1--D5 coding, evidence maturity,
Q1--Q8 appraisal, experiment-level variables, and missing/not-applicable
coding.
\\

\texttt{20\_FIGURE\_DATA}
&
Auditable source data for the manuscript figures, including PRISMA,
descriptive distributions, taxonomy results, and cross-dimensional analyses.
\\

\texttt{21\_VALIDATION}
&
Integrity checks covering corpus counts, stratification, PRISMA arithmetic,
raw-data preservation, taxonomy coverage, experiment-level denominators,
contrast admissibility, and figure consistency.
\\

\bottomrule
\end{tabular}

\vspace{2pt}
\begin{minipage}{0.97\textwidth}
\footnotesize
\textit{Note}: All \(N=289\) included publications were examined in complete
full text. The \(n=231\) original experimental studies form the denominator
for experiment-level extraction and Q1--Q8 appraisal; the remaining 58
publications remain included in corpus-level mapping and qualitative synthesis
but are not assigned experiment-level scores or frequencies.
\end{minipage}
\end{table*}
\newpage

\bibliographystyle{ieeetr}
\bibliography{sn-bibliography}

\end{document}